\documentclass[a4paper,11pt]{article}

\usepackage[tbtags]{amsmath}
\usepackage{bm}
\usepackage{amsfonts}
\usepackage{graphics}
\usepackage{graphicx}
\usepackage{psfrag}
\usepackage{eurosym}
\usepackage{pstricks}
\usepackage{array}
\usepackage{shortvrb}
\usepackage{epsf}
\usepackage{graphicx}
\usepackage{rotating}
\usepackage{float}
\usepackage{color}
\usepackage{multirow}
\usepackage{float}
\usepackage{colortbl}
\usepackage{ifpdf}
\usepackage{multicol}

\usepackage[colorlinks=true,linkcolor=blue,citecolor=blue]{hyperref}

\usepackage{xurl}
\usepackage{geometry}
\usepackage{alphalph}
\usepackage{etoolbox}

\usepackage{pdfpages}

\usepackage{caption}
\usepackage{subcaption}
\usepackage[backend=biber, style=apa, natbib]{biblatex}
\newcommand{\cblue}[1]{\textcolor{black}{#1}}

\patchcmd{\subequations}{\alph{equation}}{\alphalph{\value{equation}}}{}{}
\def \Noin {{\hskip 3pt \rm /\kern -9pt \in\hskip 1pt}}

\def \R {{\rm I\kern -2.2pt R\hskip 1pt}}

\allowdisplaybreaks

\title{Optimal day-ahead offering strategy for large producers based on market price response learning}

\author{Antonio Alc\'antara$^1$ \and Carlos Ruiz$^{2,}$\footnote{Corresponding author}}
\date{%
    $^1$Department of Statistics, University Carlos III of Madrid, Avenida de la Universidad, 30, 28911 Legan\'es, Spain. Email: \url{antalcan@est-econ.uc3m.es}\\%
    $^2$Department of Statistics and UC3M-BS Institute for Financial Big Data (IFiBiD), University Carlos III of Madrid, Avenida de la Universidad, 30, 28911 Legan\'es, Spain. Email: \url{caruizm@est-econ.uc3m.es}\\[2ex]%
    \today
}

\begin{document}

\maketitle 

\section*{Abstract}

In day-ahead electricity markets based on uniform marginal pricing, small variations in the offering and bidding curves may substantially modify the resulting market outcomes. In this work, we deal with the problem of finding the optimal offering curve for a risk-averse profit-maximizing generating company (GENCO) in a data-driven context. In particular, a large GENCO's market share may imply that her offering strategy can alter the marginal price formation, which can be used to increase profit. We tackle this problem from a novel perspective. First, we propose an optimization-based methodology to summarize each GENCO's step-wise supply curves into a subset of representative price-energy blocks. Then, the relationship between the resulting market price and the energy block offering prices is modeled through a probabilistic forecasting tool: a Distributional Neural Network, which also allows us to generate stochastic scenarios for the sensibility of the market towards the GENCO strategy via a set of linear constraints. Finally, this predictive model is embedded in the stochastic optimization model employing a constraint learning approach. Results show how allowing the GENCO to deviate from her true marginal costs renders significant changes in her profits and the marginal price of the market. Additionally, these results have also been tested in an out-of-sample validation setting, showing how this optimal offering strategy is effective in a real-world market contest.

\vspace{5mm} 
\noindent \textbf{Keywords:} OR in energy, Constraint learning, Data-driven optimization, Electricity market, Optimal pricing strategy

\section{Introduction}
\label{sec:intro}

 \subsection{Background and Aim}

Power systems worldwide are undergoing a major transformation aimed at their decarbonization while securing their reliability. One of the main players in this transition is the massive integration of renewable generating technologies from different sources. These are characterized by their low CO$_2$ emissions and negligible marginal costs, but at the same time, by a non-dispatchable nature that introduces high levels of uncertainty in the operation of electrical systems. From the market perspective, especially for those based on marginal pricing, a high share of non-dispatchable producers implies high levels of price volatility, with important economic implications for all the market players. In this regard, producers with dispatchable technologies are exposed to fast-changing market conditions which may endanger the appropriate recovery of their (usually) high investment costs. On the one hand, these technologies are needed to complement renewables and guarantee system stability through the real-time balance of supply and demand. On the other hand, if they are finally dispatched, they usually set and raise the marginal price, which further contributes to price volatility.

In this uncertain environment, Generating Companies (GENCOs) need to develop effective strategies to participate in wholesale markets to guarantee a cost-effective dispatch and a reliable operation. Hence, we focus on the problem addressed by a GENCO that participates in a day ahead electricity market based on uniform marginal pricing, and seeks to derive the daily optimal offering strategy to be submitted. Furthermore, based on historical data from a real-world system, we verify that there is indeed a capacity of this GENCO to alter the price formation of the market with its offer curve. This relationship can be approximated by a machine learning forecasting tool, and embedded in the optimization problem employed in the decision-making process. The main novelty of this approach is that it is fully data-driven, making no exogenous assumptions on the market response once the offer curves are submitted. Moreover, uncertainty is explicitly modeled by a new probabilistic based methodology which allows anticipating potential unfavorable market realizations.

\subsection{Approach}

The proposed approach is based on learning the functional relationship between a large GENCO's offer curve and the cleared marginal price, and use it to optimize her offering strategy.
To this end, we consider a two-stage decision-making process. In the first-stage, the GENCO determines the offer curve to submit to the market. In the second-stage, the market price and dispatched quantities are revealed.
Motivated by power systems where participants exhibit some degree of market power, we explicitly assume that the marginal price (and hence the dispatched quantities) is a second-stage decision variable dependent on first-stage decisions (offers), external factors, and uncertain data. This is a reasonable assumption in electricity markets based on uniform marginal pricing, in which suppliers and consumers submit their offers and bids (first-stage decisions) to the market operator. The intersection of the aggregated offering and bidding curves will generate an hourly marginal price for the market (which can be considered a second-stage decision for a player with sufficient market power). Therefore, the decisions made in the first-stage by market agents have a direct impact on the formation of the marginal price and on the profit they can obtain. This is numerically verified in this work by the analysis of real market data, which shows how the offering strategy of a large producer can impact the resulting price formation of the day ahead market.

To learn this relationship (offer curve vs market price), we make use of an extension of the recently introduced Constraint Learning (CL) methodology, an empirical approach that has been receiving attention from the scientific community \parencite{fajemisin2023optimization}. In this sense, we will employ historical data from the day-ahead market to relate historical GENCO's offers with the final marginal price cleared in the market by means of fitting a machine learning model. The structure of this model will allow, despite its non-linear nature, to transform it to a set of piece-wise linear constraints, allowing its embedding in a stochastic mixed-integer optimization (MIO) problem. \cblue{This methodology is general and fully data-driven, and requires no previous assumptions regarding the market participants behaviours, or an explicit model of the market-clearing procedure,  to characterize the functional relationship between offer curves and prices.}

\cblue{However, the classical point-predictive CL methodology employed generally in the literature does not capture the uncertainty inherent in this prediction. This can be an important drawback in applications like the one addressed in this work, as a given offer curve may result in potentially different prices due to externalities that cannot be fully characterized by the CL model. Moreover, in applications where decision makers are risk-averse, it would be needed to explicitly model or quantify this level of uncertainty.
To this end, we make use of the recent Distributional Constraint Learning (DCL) methodology developed by \textcite{alcantara2023neural}, which allows probabilistic-based scenario generation for learned variables. Hence, we seek to characterize the probability distribution of the resulting prices, conditioned to the offer curve submitted to the market, and use it to derive the GENCO's optimal market strategy.}
Note that the extension of CL methodology to account for uncertainty has already been employed to solve energy-related problems, like the one addressed by \textcite{perez2022optimal}, where a retailer had to set an optimal price for consumers. However, it is solved under a stylized and simulation-based application and making use of a simple linear regression method, with limited forecast accuracy.

Therefore, the employment of the DCL methodology allows to address the GENCO's stochastic optimization problem in a fully data-driven framework, i.e., not making any assumptions on the rest of the market's participants behavior, that solely relies on historical market data.
This is a novel approach that can be extrapolated to other applications where a complex system (market, network, organization, patient, etc.) responds to a decision maker actions, and sufficient amounts of data of these interactions are available.

\subsection{Literature Review and Contributions}
\label{sec:lit_rev}

The problem of optimal offering from a large GENCO has been widely addressed in the literature from different perspectives. Some of the most common approaches are discussed in this section.

\vspace{2mm}

\textbf{Multi-level optimization.} These optimization models usually try to combine market clearing and optimal offering by using different optimization levels. For example, in a bi-level problem, the first level could be the offering strategy formulation, while the second level could be the market clearing. Since the market-clearing can admit a linear or convex formulation, KKT optimality conditions can be used to transform the bi-level approach into a mathematical program with equilibrium constraints. This bi-level formulation was used in works such as \textcite{ruiz2009pool}, and can be easily extended to account for different sources of uncertainty \citep{beck2023survey}, through robust or stochastic optimization frameworks. In particular, \textcite{kardakos2015optimal} deals with the problem of setting an offering strategy for a virtual problem plant with a stochastic three-stage bi-level approach. In contrast, in \textcite{pandvzic2013offering}, a two-stage stochastic approach was used to establish the optimal strategy of a virtual power plant selling and purchasing energy from the day-ahead and the balancing markets seeking to maximize its profit. In \textcite{han2018offering}, the optimal strategy for a photovoltaic power plant is established by a bi-level stochastic program, dealing with the uncertainty of the competitors and its photovoltaic output. On the other hand, \textcite{han2020distributionally} employs a robust approach for a wind-storage aggregator under a bi-level optimization framework. The optimal risk-averse involvement of a virtual power plant in an electricity market is studied in \textcite{lima2022risk} by comparing both stochastic programming and robust optimization. Results suggest that, with an adequate risk parameterization, similar first-stage decisions can be obtained by both techniques.
Nevertheless, works employing multi-level optimization usually share a common limitation: they are based on strong assumptions about the offering/bidding strategies of the rest of the market competitors.

\vspace{2mm}

\textbf{Game theoretic approach.} When multiple generating companies can act as price makers, Nash Equilibrium (NE) approaches can be employed to study the offering/bidding dynamics within the market. In the context of optimal offering, an NE is a set of offers such that, given all other competitors' offers, no price-maker producer can improve her revenue by unilaterally changing her offer. Under this approach, \textcite{kannan2011strategic} developed an iterative scheme that solves an NE in a market with two price-maker producers. \textcite{naebi2020epec} employed an equilibrium problem with equilibrium constraints (EPEC) together with a diagonalization method to identify a meaningful NE to study the operation of a distribution network composed of distributed energy resources and multi-microgrids with storage devices. 
\textcite{devine2023strategic} present another EPEC model to study the interaction between several price-marking firms, with market power, and a competitive fringe, and its impact on market prices and investment decisions.
\textcite{bjorndal2023energy} propose a four-stage Stackelberg game to study how a large energy storage system (ESS) can alter the electricity price formation through its bids to further increase its profit. This effect is measured under different market designs.
More recently, in works such as the one from \textcite{du2021approximating}, reinforcement learning approaches have been combined with game-theoretical ones to improve the computational efficiency of the solution. However, as stated by \textcite{steeger2014optimal}, game theoretic models are of most interest to regulators, which seek to analyze market design or performance, on the basis of agents' rationality.

\vspace{2mm}

\textbf{Market / Competitors' offering curves learning.} As indicated, one of the challenges of the optimal offering problem is the lack of knowledge about the rivals' offering strategies. In this sense, multiple works have been developed in order to identify these strategies. For example, \textcite{ruiz2013revealing} employs inverse optimization to estimate the rival producers’ offer prices that have been marginal assuming that the GENCO has knowledge of the daily market results. 
In this regard, \textcite{allen2022using} show how inverse optimization can be used to estimate the costs of players that are involved in generalized Nash equilibrium games, where the decision of one player can impact the feasible region of the rival ones, as it is the case of firms in electricity markets.
On the other hand, \textcite{mitridati2017bayesian} make use of a Bayesian approach to unveil competitors offering curves. Despite the insightful information that can be obtained from estimating rivals' offering curves, this information is already provided a posteriori in European day-ahead markets. However, learning and, more importantly, predicting competitors' offers is of key importance for an accurate characterization of scenarios in multi-level optimal offering approaches. Finally, in \textcite{chen2019trading}, an Extreme Learning Machine is employed to find the relationship between the prosumer strategy and the obtained profits and costs in distribution grids. This is, indeed, an example of applying CL in an energy market context with an out-of-sample validation scheme like the one proposed in this work, but on a smaller scale and without taking uncertainty into account.

\vspace{2mm}

Therefore, the main contributions of this work with respect to the state of the art in optimal offering problems are the following:

\begin{itemize}
    \item[--] \textbf{General contribution:} to develop a complete data-driven optimization model for a risk-averse GENCO offering strategy, making use of real-world datasets without making assumptions regarding competitors' behaviour. This ranges from the initial treatment of the original offering curves data and training of the model that will characterize the relationships between GENCO offers and marginal price, to the out-of-sample validation of the offers derived from the stochastic problem.
    \item[--] \textbf{Specific contribution 1:} to propose an optimization-based reduction technique to summarize past realizations of the market participants' hourly offering curves (Section \ref{sec:supply_disc}). 
    \item[--] \textbf{Specific contribution 2:} to take uncertainty into account through the recently introduced Distributional Constraint Learning (DCL) methodology (Sections \ref{sec:con_learn} and \ref{sec:pred_model_sce}). 
    \item[--] \textbf{Specific contribution 3:} to show the validity of the optimal pricing strategy obtained in a real-world out-of-sample application based on the Spanish electricity market (Section \ref{sec:res_out_of_samp}).
\end{itemize}

\subsection{Article Organization}

This article is organized as follows. Section \ref{sec:genco_problem} describes the stochastic optimization problem faced by GENCO when offering her energy blocks in the day-ahead market. Section \ref{sec:methodology} details the proposed methodology, from the constraint learning background, to the optimal discretization of supply curves and the final stochastic optimization problem under the DCL approach. Then, a case study is presented in Section \ref{sec:case}, including out-of-sample testing of the optimal strategy. Finally, Section \ref{sec:conclu} draws the main conclusions of this work.

\section{GENCO's Stochastic Optimization Problem}
\label{sec:genco_problem}

In this section, we describe the optimization problem that the GENCO employs to derive her optimal offering strategy. The problem is formulated using a risk-averse two-stage stochastic approach in which the marginal price of the market is unknown but dependent on her block offers. This relationship can be explicitly modeled later employing piece-wise linearizable machine learning methods trained from historical data.

\subsection{Notation}

The notation employed to formulate the stochastic problem is described in this subsection for quick reference.

\medskip
\noindent Indices and sets:
\begin{itemize}
    \item[--] $I$: Set of energy blocks, indexed by $i$.
    \item[--] $T$: Set of hourly periods within a day, indexed by $t$.
    \item[--] $\Omega$: Set of stochastic scenarios, indexed by $\omega$.
\end{itemize}

\medskip
\noindent Variables:
\begin{itemize}
	\item[--] $P^i_{t}$: Price offered for block $i$ at time $t$.
	\item[--] $Q_{t,\omega}^{i}$: Quantity of energy produced from block $i$ at time $t$ and scenario $\omega$.
	\item[--] $u_{t,\omega}^i$: Binary decision variable indicating whether price from block $i$ at time $t$ and scenario $\omega$ is below the marginal electricity price or not.
	\item[--] $s_{\omega}$, $\eta$: Auxiliary variables for CVaR formulation.
	\item[--] $\lambda_{t,\omega} :$ Marginal market price at time $t$ and scenario $\omega$.
	\item[--] $Q^{ren}_{t} :$ Offered quantity at price zero from renewable resources at time $t$.
\end{itemize}

\medskip
\noindent Parameters:
\begin{itemize} 
    \item[--] $\theta:$ Contextual information that may have an influence on the marginal price formation.
	\item[--] $C_t^i:$ Marginal generating cost for block $i$ at time $t$.
	\item[--] $\sigma_t^i$: Allowed variability for the $i$-th block price offer with respect to its generating cost at time $t$.
	\item[--] $\pi_{\omega}:$ Probability assigned to each scenario $\omega$.
	\item[--] $\alpha$: Fraction of the profit distribution to be used in the CVaR calculation.
	\item[--] $\chi$: Weight assigned to the CVaR against the expected profit.
\end{itemize}

\subsection{Formulation}

We assume the GENCO knows in advance of sending her offers the maximum energy quantity $Q_t^{\text{Max }i}$ she can produce for each time and energy block, and its associated variable production cost $C_t^i$. In an efficient market, the producer would directly send these offers (true marginal cost of production) which can be imposed in the current model by fixing $\sigma_t^i = 0$. However, positive values of the parameter $\sigma_t^i$ will let us adjust the degree to which producers with market power can modify their offers and deviate from perfect competition.


The stochastic model is formulated as follows:

\begin{subequations}\label{eq:stochastic_model_full}
\begin{align}
\underset{\Theta}{\max} \quad &(1-\chi) \sum_{\omega \in \Omega} \pi_{\omega} \sum_{t \in T} \left[\lambda_{t,\omega} Q_t^{ren} + \sum_{i\in I}  (\lambda_{t,\omega} Q_{t,\omega}^{i} - C_t^i Q_{t,\omega}^{i}) \right] + \chi \left(\eta - \frac{1}{\alpha} \sum_{\omega \in \Omega} \pi_{\omega} s_{\omega} \right)\label{eq:stoch_of}\\
\text{s.t.}&\notag \\
  &Q_t^{ren} = Q_t^{\text{Max ren}} \quad \forall t \label{eq:sto_cons1}\\
  &u_{t,\omega}^i \in \{0,1\} \quad \forall i,t,\omega \label{eq:sto_cons2}\\
  &\lambda_{t,\omega} - P_t^i \leq u_{t,\omega}^iM  \quad \forall i,t,\omega \label{eq:sto_cons3}\\
  &P_t^i - \lambda_{t,\omega} \leq (1-u_{t,\omega}^i)M  \quad \forall i,t,\omega \label{eq:sto_cons4} \\
  &Q_{t,\omega}^{i} = u_{t,\omega}^i Q_t^{\text{Max }i} \quad \forall i,t,\omega \label{eq:sto_cons5} \\
  &\lambda_{t,\omega} = f(P_t^i, \theta) \quad \forall t,\omega \label{eq:sto_cons6} \\
  &C_t^i - \sigma_t^i \leq P_t^i \leq C_t^i + \sigma_t^i \quad \forall i,t \label{eq:sto_cons7} \\
  &0 < P_t^i \leq P_t^{i+1} \quad \forall i,t \label{eq:sto_cons8} \\
  &\eta - \sum_t \left[\lambda_{t,\omega} Q_t^{ren} + \sum_i  (\lambda_{t,\omega} Q_{t,\omega}^{i} - C_t^i Q_{t,\omega}^{i}) \right] \leq s_{\omega} \quad \forall \omega \label{eq:sto_cons9}\\
  &0 \leq s_{\omega} \quad \forall \omega \label{eq:sto_cons10}
\end{align}
\end{subequations}

\noindent where $\Theta = \{P_t^i, Q_{t,\omega}^{i}, u_{t,\omega}^i, \eta, s_{\omega}\}$ is the set of optimization variables.

The objective function (\ref{eq:stoch_of}) represents the weighted sum of the GENCO expected value and the CVaR of her profit. CVaR will be employed to measure the risk taken by the producer, which equals the expected value of $100\alpha$\% scenarios with the lowest profit. In particular, we use the linear formulation of the CVaR proposed by \textcite{rockafellar2000optimization}. The parameter $\chi \in [0,1]$ is used to model the risk aversion level of the producer. Thus, when $\chi$ is equal to zero the producer acts as a risk-neutral decision-maker. On the other hand, when $\chi$ is equal to one, the producer can be considered risk-averse, and his decisions will focus on improving the left tail of the profit distribution. The expected profit is computed as the sum of the profits over the set of scenarios multiplied by their respective probabilities $\pi_{\omega}$. The profit is composed of the revenues of the dispatched energy blocks, paid at a marginal price $\lambda_{t,\omega}$, minus their production costs.

Constraint (\ref{eq:sto_cons1}) establishes that the produced renewable energy $Q_t^{ren}$ at time $t$ equals the estimated one in order to send block offers to the pool $Q_t^{\text{Max ren}}$. Constraints (\ref{eq:sto_cons2})-(\ref{eq:sto_cons4}) model whether the price requested for one block $P_t^i$ is lower than the marginal price $\lambda_{t,\omega}$, or not. In particular, (\ref{eq:sto_cons2}) assigns variable $u_{t,\omega}^i$ a binary domain while (\ref{eq:sto_cons3}) and (\ref{eq:sto_cons4}) are employed through the big-M method in order to assign $u_{t,\omega}^i$ a value of one if the price of the block $i$ is under the marginal price, and zero otherwise.

With this assigned value of $u_{t,\omega}^i$, equation (\ref{eq:sto_cons5}) will establish the dispatched energy. That is, $Q_{t,\omega}^{i}$ will be equal to the estimated energy that can be produced $Q_t^{\text{Max }i}$ for the block $i$ if the price requested for that block is under the marginal price. On the other hand, the value of $Q_{t,\omega}^{i}$ will be zero if the price of that block is over the marginal price. That is, it is not a dispatched energy block. 

Equation (\ref{eq:sto_cons6}) establishes that the stochastic marginal price can be estimated through a model $f(\cdot)$, and therefore is dependent on the offering block prices $P_t^i$, and external contextual information $\theta$. This contextual data may include information that usually has an influence on the marginal price of the market, such as renewable energy production or demand. The specific methodology to generate each scenario realization of the price $\lambda_{t,\omega}$ in (\ref{eq:sto_cons6}) is introduced in Section \ref{sec:pred_model_sce}.

Constraint (\ref{eq:sto_cons7}) determines the producer flexibility of setting a block price $P_t^i$ above or below its true marginal cost $C_t^i$, as a function of parameter $\sigma_t^i$. In particular, the impact of $\sigma_t^i$ will be used to study how the GENCO can increase her profit by modifying the offering block prices. Equation (\ref{eq:sto_cons8}) ensures an increasing offering curve, a condition required in most electricity markets (see Figure \ref{fig:curve_discr}). Finally, constraints (\ref{eq:sto_cons9}) and (\ref{eq:sto_cons10}) will be employed in order to characterize the CVaR. In particular, the value of $\eta$ would be equal to the Value at Risk at the optimal solution to the problem (\ref{eq:stochastic_model_full}).

One of the advantages of this formulation is that it can be easily transformed into a mixed-integer linear problem, as the only non-linear term is the product between two variables: $\lambda_{t,\omega} Q_{t,\omega}^{i}$, which according to (\ref{eq:sto_cons5}) is equivalent to $\lambda_{t,\omega}u_{t,\omega}^i Q_t^{\text{Max }i}$. The product $ \lambda_{t,\omega}u_{t,\omega}^i$ involves a continuous variable and a binary one, which can be linearized without approximation (big-M approach). It is important to notice that equation (\ref{eq:sto_cons5}) relies on the assumption that all energy is dispatched when the GENCO offer is under the marginal price. This approximation, which improves with the number of blocks considered, excludes those cases where the strategic producer marginal block can be partially dispatched due to technical or economic reasons. However, it is needed to simplify bilinear terms in the objective and keep the optimization problem tractable.

\section{Methodology}
\label{sec:methodology}

The following sections will describe the developed methodology in order to learn the functional relationship of GENCO's offers and the marginal price formation. First, Section \ref{sec:supply_disc} will propose an optimization-based approach (\ref{eq:discr_model}) to obtain insights from GENCO's offers in the form of discrete offer blocks. Then, Section \ref{sec:con_learn} will introduce the DCL methodology, employed to learn the relationship between the offered energy blocks and the marginal market price distribution. Finally, Section \ref{sec:pred_model_sce} will merge all the information, explaining how can the predictive model be embedded within the optimization problem, and obtaining the final stochastic optimization model for GENCO's problem (\ref{eq:stochastic_model_full2}).

\subsection{Optimal discretization of supply curves and marginal price formation}
\label{sec:supply_disc}

As has been mentioned before, the GENCO's optimization problem is based on deciding the prices offered to the market for her blocks of energy while predicting their possible impact on the marginal market price.

In general, in day-ahead electricity markets based on uniform marginal pricing, data from the hourly supply curve of each GENCO is composed of production units' power blocks with their corresponding offering prices, i.e., prices that each unit is willing to accept to produce that amount of power for one hour. In a fair, transparent, and audited market, this price must reflect the marginal generating cost of that production unit.

\begin{figure}[ht]
    \centering
    \includegraphics[width=0.65\textwidth]{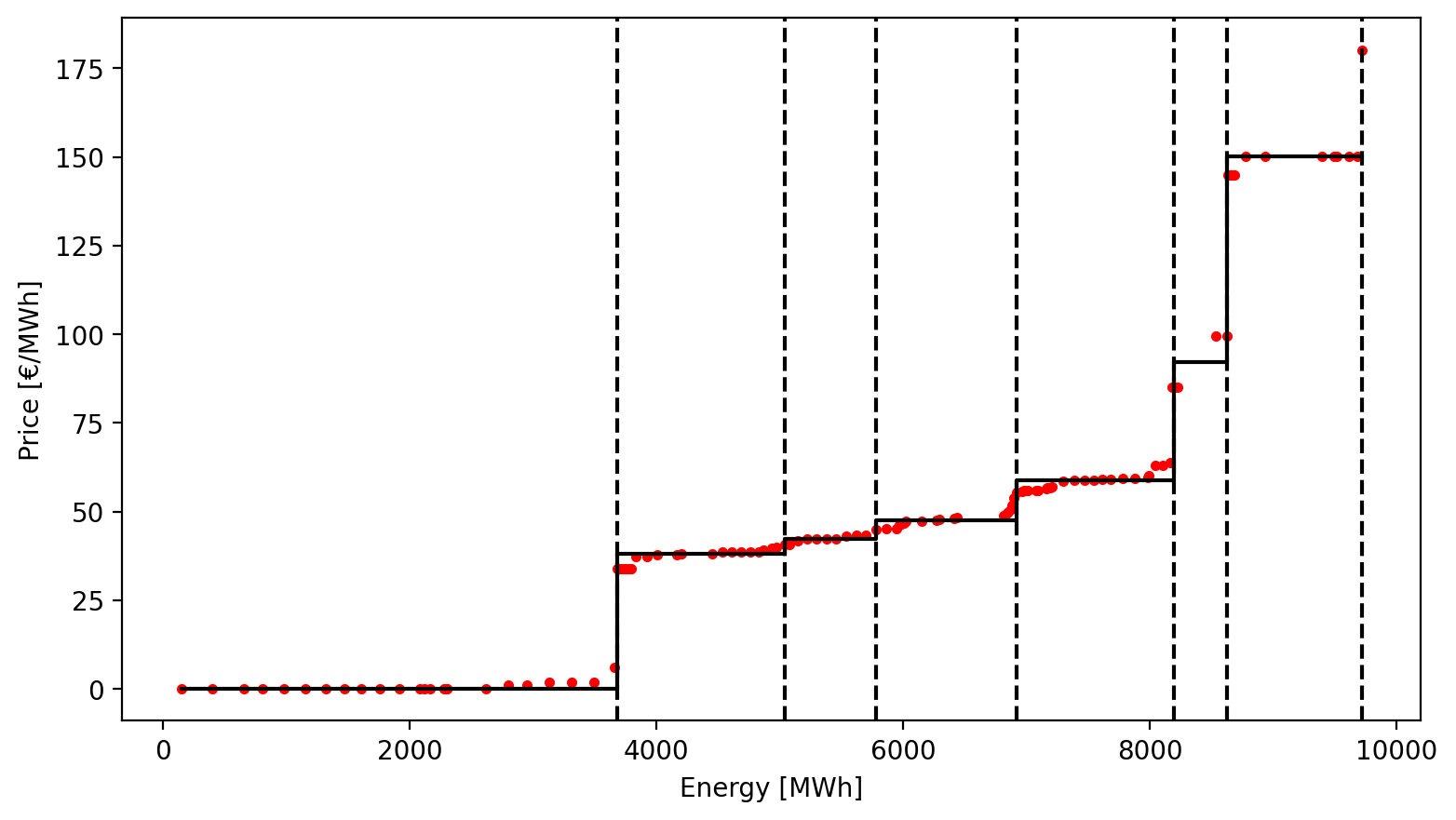}
    \caption{Supply curve discretization example for June 3rd 2017 at 15:00. Source: OMIE}
    \label{fig:curve_discr}
\end{figure}

To build the GENCOS' aggregated supply curve for each hour, production blocks are ordered by their price, from the lowest to the largest. Regarding the energy quantity, a cumulative sum is performed so that each price faces the cumulative quantity of energy of those blocks with prices below. This renders an increasing step-wise curve. An illustrative example can be seen in Figure \ref{fig:curve_discr}, where for a given hour and particular GENCO, red points represent her production units with their price and cumulative quantity within the supply curve.

For tractability of the subsequent piece-wise linearizable predictive model (which will be employed under the DCL methodology), we propose to summarize first each aggregated supply curve into a smaller number of blocks by using an optimization approach (\ref{eq:discr_model}). The aim is to obtain a minimum-error-based discretization of GENCO's hourly supply curve, apart from setting clear decision variables for the optimization problem (\ref{eq:stochastic_model_full}).

\begin{subequations}\label{eq:discr_model}
\begin{align}
\underset{C_{b},\delta_b}{\min} & \quad \sum_{b=1}^{B}|C_b-P_b^{R}| q_b^R \label{eq:discr_model_OF}\\
\text{s.t.:}&\notag\\
  &0 \leq C_b-C_{b-1} \leq \delta_b M^P \quad b=2,\dots,B \label{eq:discr_model_cons1}\\
  &0 \leq C_b-C_{0} \leq \delta_1 M^P \quad b=1 \label{eq:discr_model_cons2}\\
  &\sum_{b=1}^B \delta_b=|I|-1 \label{eq:discr_model_cons3}\\
  &\delta_{b} \in \{0,1\} \quad b=1,\dots,B \label{eq:discr_model_cons4}
\end{align}
\end{subequations}

In this problem, we seek to obtain $|I|$ grouped power blocks from a total of $B$ original ones. $C_b$ is the energy price, $P_b^R$ is the real price observed, and $q_b^R$ is the amount of energy of each block $b$. The objective function of the model (\ref{eq:discr_model_OF}) aims to minimize the absolute distance between the real prices $P_b^{R}$ and the optimized ones $C_b$, weighted by the energy quantity of the block. Constraints (\ref{eq:discr_model_cons1}) and (\ref{eq:discr_model_cons2}) allow one to assign the same price $C_b$ to all the values that belong to the same grouped block. The sum of $\delta_b$ in constraint (\ref{eq:discr_model_cons3}) ensures that $|I|-1$ cuts are obtained, and therefore $|I|$ grouped blocks, since $\delta_b$ is defined as a binary variable (\ref{eq:discr_model_cons4}).

The solution to this problem will let us know the different prices $C_b$ and positions $\delta_b$ where a step in the curve can be set to obtain an optimal cut and, therefore, an optimal grouped block. As an example, let us assume that we want to get $|I|=7$ grouped blocks from the curve (red dots) presented in Figure \ref{fig:curve_discr}. The optimization problem will render the solid black step-wise curve as a solution. In this way, the dashed vertical lines represent the different cuts, and solid horizontal lines the energy price $C_b$ associated with each grouped block. Note that, for simplicity, an extra vertical line has been added in the last production unit to set the last block.

\cblue{The selection of $|I|$ responds to a trade-off between solving times and fitting accuracy: the higher $|I|$, the more accurate approximation of the supply curve (\ref{eq:discr_model_OF}), but with a higher problem complexity as we increase the number of binary variables and constraints. Furthermore, a high value of $|I|$ may lead to non-meaningful solutions. This problem happens when most price values $P_b^R$ are not unique, which makes it not possible to obtain unique grouped blocks $C_b$ following the formulation of the optimization problem.} 

As can be seen, this solution will appropriately summarize the information within the original supply curve, as it results in a finer discretization for those parts of the curve with higher price variations. This will also lead to obtain a historical dataset of summarized offering block prices and energy quantities, which is needed for the later training of the machine learning model under the DCL methodology.

This optimal discretization problem has been already addressed in the literature. For example, \textcite{ruan2020data} modeled the problem in a similar sense for residual demand curves. However, their objective function was quadratic, and a preprocessing procedure was needed before solving the actual problem.

Once we obtain the optimal discretization of the GENCO supply curve, the same procedure can be applied to the rest of the offers on the market, that is, the supply curves of the competitors. Considering both the GENCO's and competitors' supply curves, and ordering all the blocks by their price, the cumulative energy quantity can be computed to build the aggregated discretized market supply curve.

Note that it is also possible to achieve a summarized curve by directly discretizing the original aggregated supply curve. However, splitting GENCO's units and the ones from the competitors will allow us to know the weight of the producer within the market and to get insights into how its pricing strategy can modify the market outcomes. 

Nevertheless, to set a marginal price, a demand curve is also needed. For the sake of simplicity, we will assume an inelastic demand curve, which will cut the supply curve and establish the price producers will be paid for each MWh of energy offered below this price.

\begin{figure}[ht]
    \centering
    \includegraphics[width=0.65\textwidth]{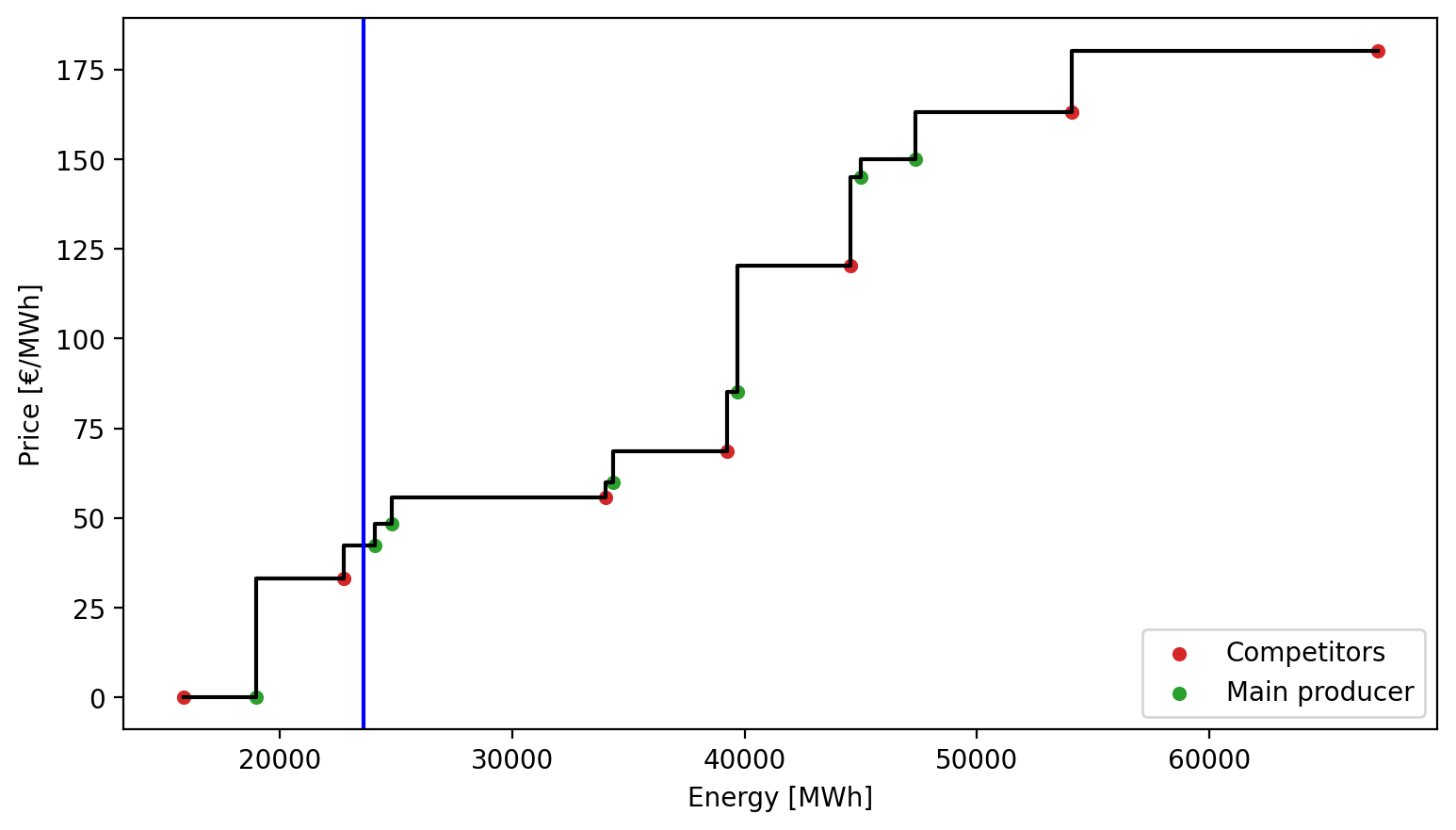}
    \caption{Example of marginal price formation on the discretized market}
    \label{fig:disc_market}
\end{figure}

\cblue{An example of this procedure is shown in Figure \ref{fig:disc_market}, where the increasing step-wise curve represents the aggregated market supply curve, and the vertical line represents the inelastic demand. In this curve, steps ending in a green dot identify the blocks offered by the GENCO (main producer), while steps ending in a red dot stand for competitors' blocks. We can see a total dispatched energy of 24000 MWh at a price of 42 \euro/MWh. Notice how the marginal price is set by a block from the main producer in this case. This motivates the study of how small variations in the offered price can directly change the marginal market price.}

\begin{figure}
    \centering
    \begin{subfigure}{0.99\textwidth}
    \centering
    \includegraphics[width=\textwidth]{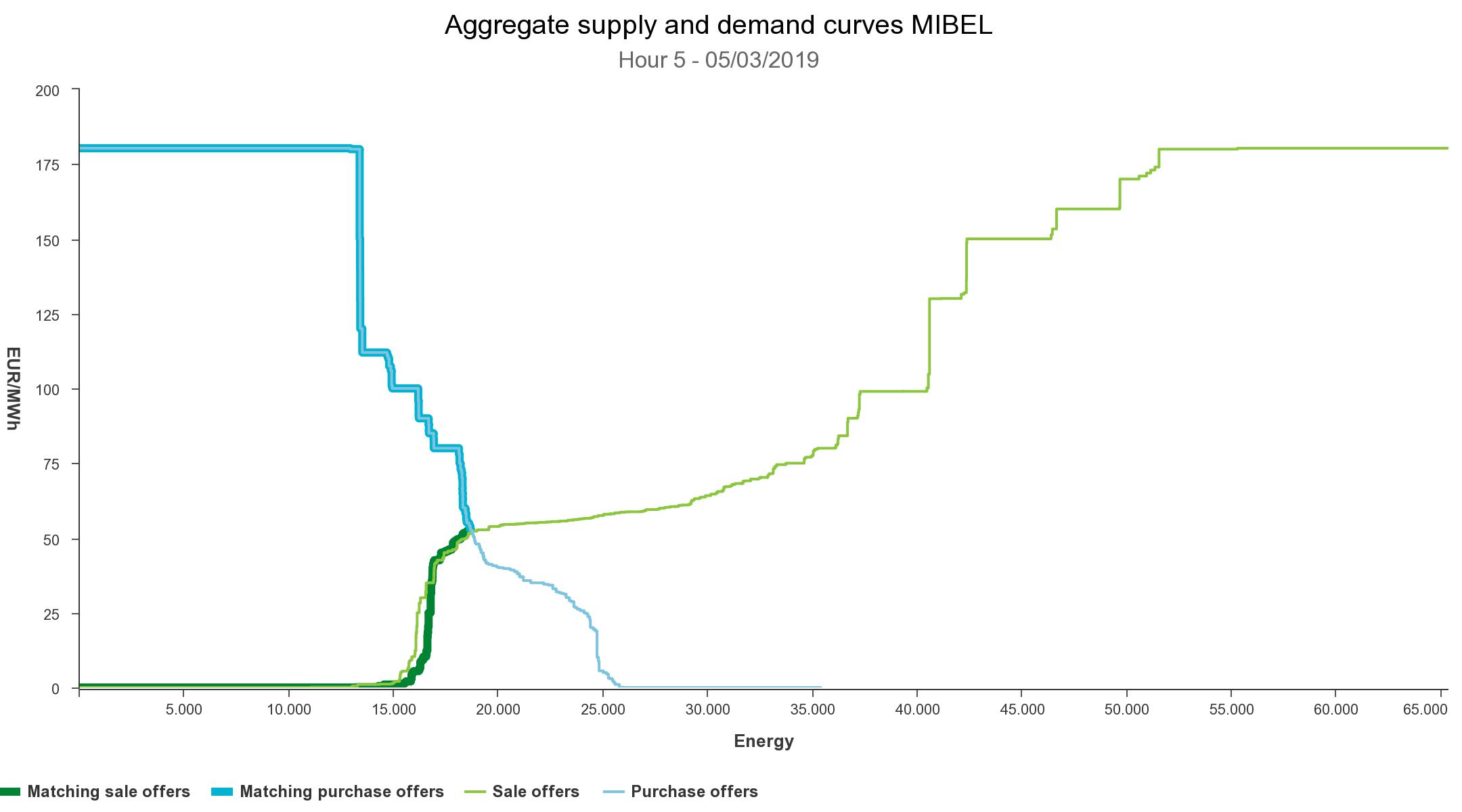}
    \end{subfigure}
    \hfill
    \begin{subfigure}{0.99\textwidth}
    \centering
    \includegraphics[width=\textwidth]{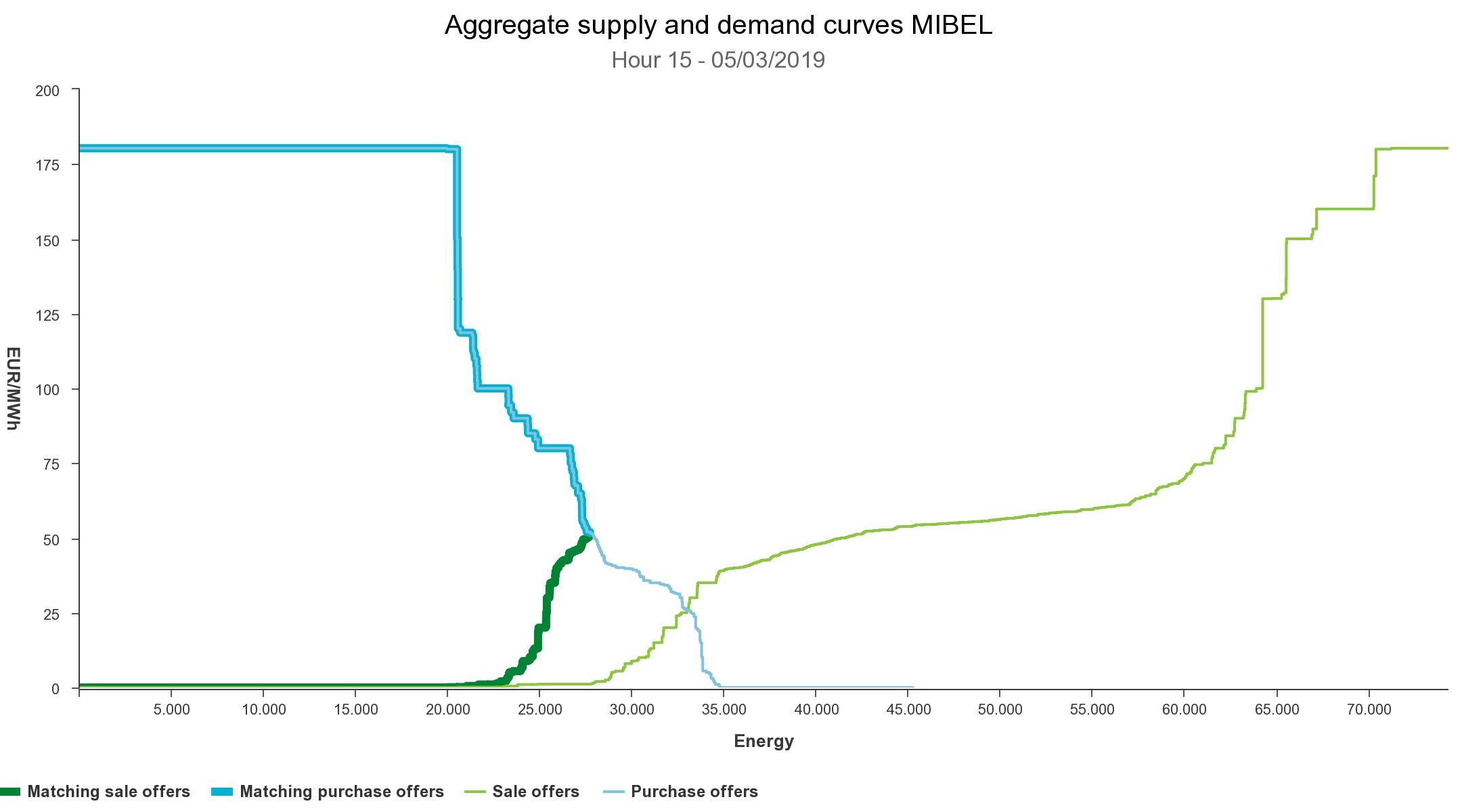}
    \end{subfigure}
    \caption{Marginal price formation in two different hours in the MIBEL}
    \label{fig:omie_curves}
\end{figure}

However, in real day-ahead electricity markets, the price resulting from the cut of the submitted demand and supply curves is not necessarily the resulting market marginal price (e.g., Spain and Portugal). In fact, it is common to observe that the final supply curve from the market suffers changes (withdrawal of some operation units) due to the system operator's ex-pot verification of technical constraints. We can take as an example the market curves presented in Figure \ref{fig:omie_curves}. There, two different hourly marginal price formation procedures are shown within the same day in the Iberian Electricity Market \parencite{MIBEL}. The differences between the sale offers and the matching (dispatched) sale offers can be easily noticed at hour 15.

Therefore, when analyzing historical market data, it is important to consider this fact by statistically characterizing the differences between the matching supply curve and the offered one. This will allow us to validate the optimal strategy in real life (out-of-sample). That is, we can set the optimal blocks with real data and use competitors' blocks, jointly with the demand and the approximated displacement of the supply curve to study the differences in profit according to the producer risk aversion and other relevant features.

\subsection{Distributional Constraint Learning Framework}
\label{sec:con_learn}

As we have seen from the literature review exposed in Section \ref{sec:lit_rev}, most approaches for setting optimal offerings are based on large assumptions or theoretical modeling, far from the real knowledge we can obtain from historical observation of the market. In this sense, our aim is to develop a fully data-driven methodology that allows us to estimate the marginal price (crucial in the optimal offering) based on the GENCO offers. This relationship can be embedded as sets of constraints under the so-called constraint learning (CL) framework \parencite{fajemisin2023optimization}.

A generalization of the CL framework can be seen in the optimization problem (\ref{eq:cl_opt}), which aims to minimize the expected value of the cost function $c(x,y;\theta,\xi): \mathbb{R}^{d_x} \rightarrow \mathbb{R}$ (\ref{eq:cl_OB}). This cost depends not only on the decision variables $x \subset \mathbb{R}^{d_x}$ and $y \subset \mathbb{R}^{d_y}$ (equivalent to $P^i_t$ and $\lambda_{t,\omega}$ in our application (\ref{eq:stochastic_model_full}), respectively), but also on external contextual information $\theta \subset \mathbb{R}^{d_{\theta}}$ and uncertainty $\xi \subset \mathbb{R}^{d_{\xi}}$. Note that additional constraints could be applied through known functions $g(\cdot)$ on $x$ and $y$, as stated in the constraint (\ref{eq:cl_cons1}).
\begin{subequations}\label{eq:cl_opt}
\begin{align}
\underset{x, y}{\min} \;& \mathbb{E}\left[ c(x,y;\theta,\xi) \right] \label{eq:cl_OB}\\
\text{s.t.}&\notag \\
  &g(x,y)\leq 0  \label{eq:cl_cons1}\\
  &y = f^{\mathcal{D}}(x;\theta,\xi) \label{eq:cl_cons2} 
\end{align}
\end{subequations}

The learned constraint is established in (\ref{eq:cl_cons2}). The relationship between $y$ and $x$ (and contextual information $\theta$ and uncertainty $\xi$) is modeled using a piece-wise linearizable prediction model $f^{\mathcal{D}}$, which is trained using a data set $\mathcal{D} = \{x_i,y_i,\theta_i \}_{i=1}^N$.

Recently, the Distributional Constraint Learning (DCL) methodology was developed in order to take into account statistical uncertainty around the learned decision variable $y$ \parencite{alcantara2023neural}. This approach makes the CL framework suitable for its implementation in stochastic optimization problems. 

\cblue{An example of the DCL methodology for the GENCO's strategic offering problem can be seen in (\ref{eq:dcl}). Now, we can consider the offering strategy of the GENCO $P^i_t$ as $x$, while the price-response from the market $\lambda_t$ corresponds to $y$. Furthermore, contextual information $\theta$ can include calendar information, lags of the day-ahead price, temperature, demand, or renewable energy forecasts, among others. Differently from the general CL methodology, now the output from the piece-wise linearizable machine learning model will be composed of an estimate of the mean and the standard deviation for the distribution of the day-ahead price $\lambda_t$ (\ref{eq:dcl_modelcons}).}
\begin{subequations}\label{eq:dcl}
\begin{align}
\underset{P_t, \tilde{\lambda}_t}{\min} \;& \mathbb{E}_{\tilde{\lambda}_t \sim \mathcal{Y}(\hat{\mu}_{\lambda_t}, \hat{\sigma}_{\lambda_t})}\left[ c(P_t,\tilde{\lambda}_t;\theta,\xi) \right] \label{eq:dcl_obj}\\
\text{s.t.}&\notag \\
  &g(P_t,\tilde{\lambda}_t) \leq 0  \label{eq:dcl_othercons}\\
  &\hat{\mu}_{\lambda_t}, \hat{\sigma}_{\lambda_t} = f^{\mathcal{D}}(P_t,\theta,\xi) \label{eq:dcl_modelcons}
\end{align}
\end{subequations}

\cblue{With this formulation we establish the functional relationship between GENCO's offers (and additional contextual information) and the resulting day-ahead price distribution $\mathcal{Y}(\hat{\mu}_{\lambda_t}, \hat{\sigma}_{\lambda_t})$. In this way, the relationship (\ref{eq:dcl_modelcons}) will provide enough insights into how the offering decisions could impact the day-ahead price at solving time. A neural network will be employed as the predictive model $f^{\mathcal{D}}$, whose structure and peculiarities for generating decision-dependent scenarios are exposed in Section \ref{sec:pred_model_sce}.}

\subsection{Marginal price predictive model and scenario generation}
\label{sec:pred_model_sce}

To strategically derive the GENCO's optimal offered prices within her supply curve (one for each block) is necessary to find a function accurate enough to predict the response of the marginal electricity price. That is, assuming that the strategic GENCO can alter the market marginal price, we will characterize it as a function of its price block offers and other external covariates, such as different marginal price lags, the expected demand, or levels of renewable energy production, for example.

Let us assume we have a dataset of $N$ observations of the type $\mathcal{D} = \{(\lambda_1, \mathbf{X}_1, \bm{\theta}_1), \dots,$ $(\lambda_N, \mathbf{X}_N, \bm{\theta}_N) \}$, where $\lambda \in \mathbb{R}$ represents the marginal price, $\mathbf{X} = \{P^1, \dots, P^{d_x}\}$ are the block price offers done by the GENCO, and $\bm{\theta} = \{\theta^1, \dots, \theta^{d_{\theta}}\}$ are the values for the rest of auxiliary variables (covariates) the marginal price depends on, known at the time of the decision making.

In this way, it is possible to characterize the marginal price $\lambda$ as a piece-wise linear function of $\mathbf{X}$ and $\bm{\theta}$, allowing its inclusion in optimization problem (\ref{eq:stochastic_model_full}) as a set of constraints. Following the DCL methodology, a Distributional Neural Network (DNN) can be employed for this task \parencite{nix1994estimating}. This DNN is fitted by minimizing the Gaussian negative log-likelihood, that is, we assume the estimation for the dependent variable is a random variable that follows a normal distribution. \cblue{The use of the Gaussian distribution has been widely used in electricity price forecasting tasks, see for example \cite{contreras2003arima, islyaev2015electricity, brusaferri2019bayesian}; or \cite{grothe2023point}.}

\cblue{Within the context of our framework, we posit the Gaussian negative log-likelihood loss as the proper choice. Our objective is not only generating day-ahead price predictions conditioned by GENCO's supply curves; rather, we aim to account for the inherent uncertainty in marginal prices enveloping our decisions. The Gaussian loss owns the capacity to simultaneously account for point and variance estimations derived from the DNN. This improves the understanding of prediction reliability, while also enabling risk-averse optimization.}

The structure of this DNN is similar to a classical fully-connected neural network. It is made up of an input layer, a single or multiple hidden layers, and an output layer. Each of these layers is composed at the same time of neurons, where matrix operations and non-linear functions are applied. In general, for a given hidden layer $l \in L$ of the DNN with nodes $N^l$, the value of a node $i \in N^l$, denoted as $v_i^l$, is calculated using the weighted sum of the node values of the previous layer, followed by a non-linear activation function $h(\cdot)$. This value is given as:
\begin{equation}
\label{eq:output_1ayer}  
 v_i^{l}=h\left(b_i^l + \sum_{j\in N^{l-1}} w_{ij}^{l} v_j^{l-1} \right)
\end{equation}
\noindent where $b_i^l$ represents the bias term, and $w_{ij}^{l}$ the weights for node $i$ in layer $l$. This computation will continue in the following layers, taking as input the output of the previous layers, until the final one is reached.

The particularity of the DNN lies in its output: it throws an estimate of the mean and standard deviation of the dependent variable distribution, assuming its normality as mentioned above. From these conditional distribution parameters, we will be able to generate scenarios for the marginal price $\lambda_{t,\omega}$ stated in constraint (\ref{eq:sto_cons6}), as we will show later in this section.

For the embedding of the DNN within the optimization problem as a set of constraints, we have to take into account mainly the non-linear activation function $h(\cdot)$. Some works like \cite{anderson2020strong} have studied how to implement trained neural networks as mixed-integer formulations. One of the most employed options is to train the neural network using the Rectified Linear Unit (ReLU) as the non-linear activation function. This function is composed of a max operator, that can be easily transformed to a set of piece-wise linear constraints \citep{anderson2020strong}
%
%

Once the predictive model has been fitted and embedded within the stochastic optimization problem, adding one simple constraint will allow us to generate scenarios for $\lambda_{t,\omega}$ from the mean and standard deviation estimate of its distribution. In this way, we can turn the optimization problem (\ref{eq:stochastic_model_full}) into the following one:
\begin{subequations}\label{eq:stochastic_model_full2}
\begin{align}
\underset{\Theta}{\max} \quad &(1-\chi) \sum_{\omega \in \Omega} \pi_{\omega} \sum_{t \in T} \left[\lambda_{t,\omega} Q_t^{ren} + \sum_{i\in I}  (\lambda_{t,\omega} Q_{t,\omega}^{i} - C_t^i Q_{t,\omega}^{i}) \right] + \chi \left(\eta - \frac{1}{\alpha} \sum_{\omega \in \Omega} \pi_{\omega} s_{\omega} \right)\label{eq:stoch_of2}\\
\text{s.t.}&\notag \\
  &(\ref{eq:sto_cons1}) - (\ref{eq:sto_cons5})\\
  &\hat{\mu}_{\lambda_{t}}, \hat{\sigma}_{\lambda_{t}} = f(P_t^i, \theta) \quad \forall t \label{eq:sto_cons_model} \\
  &\lambda_{t, \omega} = z_{\omega} \times \hat{\sigma}_{\lambda_{t}} + \hat{\mu}_{\lambda_{t}}, \quad \forall t,\omega \label{eq:dcl_yrandom} \\
  &(\ref{eq:sto_cons7}) - (\ref{eq:sto_cons10})
\end{align}
\end{subequations}

Now, in our optimization problem, constraint (\ref{eq:sto_cons_model}) represents the DCL framework. As can be seen, the DNN $f(\cdot)$ will establish the relationship between our offering prices $P_t^i$ (and contextual information $\theta$ that has an influence on the marginal price of the market, such as renewable energy production or demand) with the parameters of the conditional distribution of the hourly marginal price $\lambda_t$, i.e., $\hat{\mu}_{\lambda_{t}}$ and $\hat{\sigma}_{\lambda_{t}}$. On the other hand, constraint (\ref{eq:dcl_yrandom}) allows us to generate random scenarios for the marginal price by reconstructing a normal distribution using random values from a standard normal $z_{\omega}$ sampled outside the optimization problem. The joint fulfillment of constraints (\ref{eq:sto_cons_model}) and (\ref{eq:dcl_yrandom}) allows the generation of scenarios for the marginal price directly dependent on the GENCO's decisions. Therefore, the producer will try to alter the marginal price distribution to maximize her profit distribution.

The main advantage of the developed methodology is to avoid assuming the competitor's offering strategy, as the DNN has the generalization capacity for complex functional relationships. Therefore, we can solve GENCO's problem in a purely data-driven way. On the other side, the main disadvantages may come with a possible lack of some explanatory variables or big changes in the price formation process due to external factors, which will lead to wait for more data in order to retrain the model. Furthermore, it is important to have enough data to train the model, as optimizing over pre-trained machine learning models on low-density areas could raise generalization problems \citep{maragno2023mixed}.


\section{Case study}
\label{sec:case}

The main objective of this case study is to analyze the optimal bidding strategy of a large GENCO in a data-driven context. For this purpose, real-world data will be employed during the study, jointly with an out-of-sample validation, to show how price variations in GENCO's offered energy blocks can modify the marginal market price and make her profit increase.

\subsection{The dataset}

In this case study, we will focus on a large GENCO within the Spanish electricity market. This GENCO produces around 25\% of the total energy in the system. As exposed in Section \ref{sec:methodology}, we will employ different types of data to characterize the problem.

Firstly, in order to obtain discretized energy blocks and compute a marginal price with the inelastic demand, the hourly supply curves for each GENCO in the day-ahead market are necessary. This information is open-access and provided by the designated electricity market operator for the Iberian Peninsula, OMIE \parencite{OMIE}. Six years of data have been collected on an hourly frequency, from June 2017 to June 2023. This data includes day-ahead hourly supply curves (similar to Figure \ref{fig:curve_discr}) for the production units from the large GENCO as well as from the competitors. That is, more than 100 million observations were processed. Hourly data from June 1st, 2017 to May 31st, 2023 will be mainly employed to train our linear model to predict the day-ahead marginal price. June 2023 will be used as a test, applying an out-of-sample validation approach.

It is important to notice that, in the selected time range, the electricity price went from a relatively stable period between 2017 and 2019, to lower prices with COVID lockdowns, and to huge increments with the gas and oil price higher prices and inflation.

Regarding the covariates employed in the predictive model (price lags, demand, solar, and wind energy forecasts with their respective lags), they are obtained through the information system of Red El\'ectrica Española (the Spanish system operator), ESIOS \parencite{ESIOS}. All this information is assumed to be known in advance at the decision-making time.

\subsection{Applied methodology}

The methodology for solving the GENCO's optimization problem under the DCL approach is based on three steps. First, we discretize the market offering curves. Then, we train the DNN model to learn the relationship between the marginal price, GENCO' offers, and the rest of the contextual variables. Lastly, we embed the fitted model into the optimization problem, and generate decision-dependent scenarios at decision-making time.

\subsubsection{Market dicretization}

The first step in the methodology process is to optimally discretize the offering supply curves. As exposed in Section \ref{sec:methodology}, we will consider a specific supply curve discretization for the GENCO and another one for the rest of the competitors. This allows better characterizing the GENCO's curve and gaining technological insights, which will be useful in designing the marginal price prediction model.

All the hourly supply curves will be discretized into 7 different blocks, both for the main producer and for the competitors. For each block, we assume its energy quantity and price represent how much quantity can be produced and at what cost. We consider 7 blocks as a reasonable number of blocks to capture the main functional properties of the supply curves, while not increasing in excess the dimensionality of the mixed-integer linear problem (\ref{eq:stochastic_model_full}). The first block will always represent the amount of estimated renewable production (at zero cost), and the last two blocks, high-cost generating technologies that, as the historical data reflects, are never marginal. An example of this characterization was shown in Figure \ref{fig:curve_discr}. 

Once the discretized market supply curve is obtained, we approximate the hourly aggregated demand curve by an inelastic one. We get this value from the estimated hourly demand provided by ESIOS. This inelastic demand, jointly with the aggregated supply curve will render an intersection point, that sets a first marginal price with its corresponding total dispatched energy (Figure \ref{fig:disc_market}). 

However, we know that this first intersection can be far from the resulting market price value, as technical restrictions come into play. These change the shape of the supply curve, creating a gap between the intersection of the curves and the resulting hourly marginal price (Figure \ref{fig:omie_curves}). 

This is a phenomenon that we can not precisely quantify without a detailed physical description of the electricity system, but that must be considered somehow by the GENCO, as it may modify the resulting market outcomes. Nevertheless, by using historical data, we have statistically characterized this difference, and used it to replicate the resulting supply curve including these technical corrections. The difference in energy quantities will allow us to displace our discretized supply curve and give a more accurate estimation of the market marginal price. After this displacement is applied, we compute our new intersection, obtaining the final market marginal price and the dispatched energy.

For fair out-of-sample model validation, the real hourly displacement due to the technical restrictions cannot be known in advance by the GENCO. Thus, as a proxy, we will employ the mean displacement of the last two months, grouped by hourly periods. We will see that this is an effective strategy to characterize this phenomenon.

\subsubsection{Marginal price predictive model}

Now that hourly marginal prices from the discretized supply curves have been computed, we seek the adequate predictive model (\ref{eq:sto_cons_model}), i.e., to estimate the marginal price as a function of the offering prices and several covariates, to be embedded within the stochastic optimization problem (\ref{eq:stochastic_model_full2}). 

As stated in Section \ref{sec:methodology}, a DNN will be used under the DCL methodology in order to generate scenarios from the hourly day-ahead price distribution, conditioned on the GENCO's offers. For every NN-based model, we must choose the size and shape of its structure. In our case, we will employ 1 hidden layer with 100 neurons in the DNN. This self-contained structure size has been chosen to deal with the trade-off between network complexity and performance \parencite{alcantara2023neural}.

For this DNN model, the following predictors have been used:
\begin{itemize}
	\item[--] Quantity of renewable energy offered by the large GENCO power producer.
	\item[--] Block prices (6 decision variables) offered by the GENCO. These decision variables include all the block prices except the one related to the renewable energy, which is offered at zero price.
	\item[--] Demand estimation for the respective hour, and 7 lags (24, 48, ..., 168 hours), 24 hour rolling mean, maximum and minimum.
	\item[--] Wind power estimation for the respective hour, and 7 lags (24, 48, ..., 168 hours), 24 hour rolling mean, maximum and minimum.
	\item[--] Solar power estimation for the respective hour, and 7 lags (24, 48, ..., 168 hours), 24 hour rolling mean, maximum and minimum.
	\item[--] 7 marginal price lags (24, 48, ..., 168 hours).
	\item[--] Calendar covariates: dummies for day of the week and month, binary variable for holidays COVID period and time when the maximum limit for price offering was increased from 180\euro{} to 3000\euro{}.
\end{itemize}

Therefore, the dataset will be formed by a total of 70 different predictors for the hourly market marginal price model, being trained in the six-year hourly period stated previously.

Notice that the predictive performance of the DNN could be improved if more predictors were available. For instance, given the importance of hydropower in the Spanish market, knowing in advance the hydro production would be beneficial for our model. However, we have focused on external regressors that are relatively easier to forecast, are well-known in advance, and are available on the ESIOS platform. Solar and wind energies are non-dispatchable sources, while hydro-power is dispatchable and therefore, subject to the producer's decision. The goal of the DNN is to characterize this uncertain rival behavior through the probabilistic characterization of the impact of the GENCO offers on the resulting market prices.

The trained DNN with the mentioned covariates can output a probabilistic forecast of the day-ahead price in the form of an estimate for its mean and standard deviation parameters assuming a normal distribution. First, to make a simple evaluation of the DNN predictive performance, we will focus on whether the values generated from the estimated distribution can cover the real value of the day-ahead price. Concerning this coverage of the distributions on the test days, we can say that by generating 300 samples from the estimated distributions, around 97\% of the hourly prices are covered. Notice that the DNN outputs probabilistic estimates and not point predictions.

\begin{figure}
    \centering
    \includegraphics[width=\textwidth]{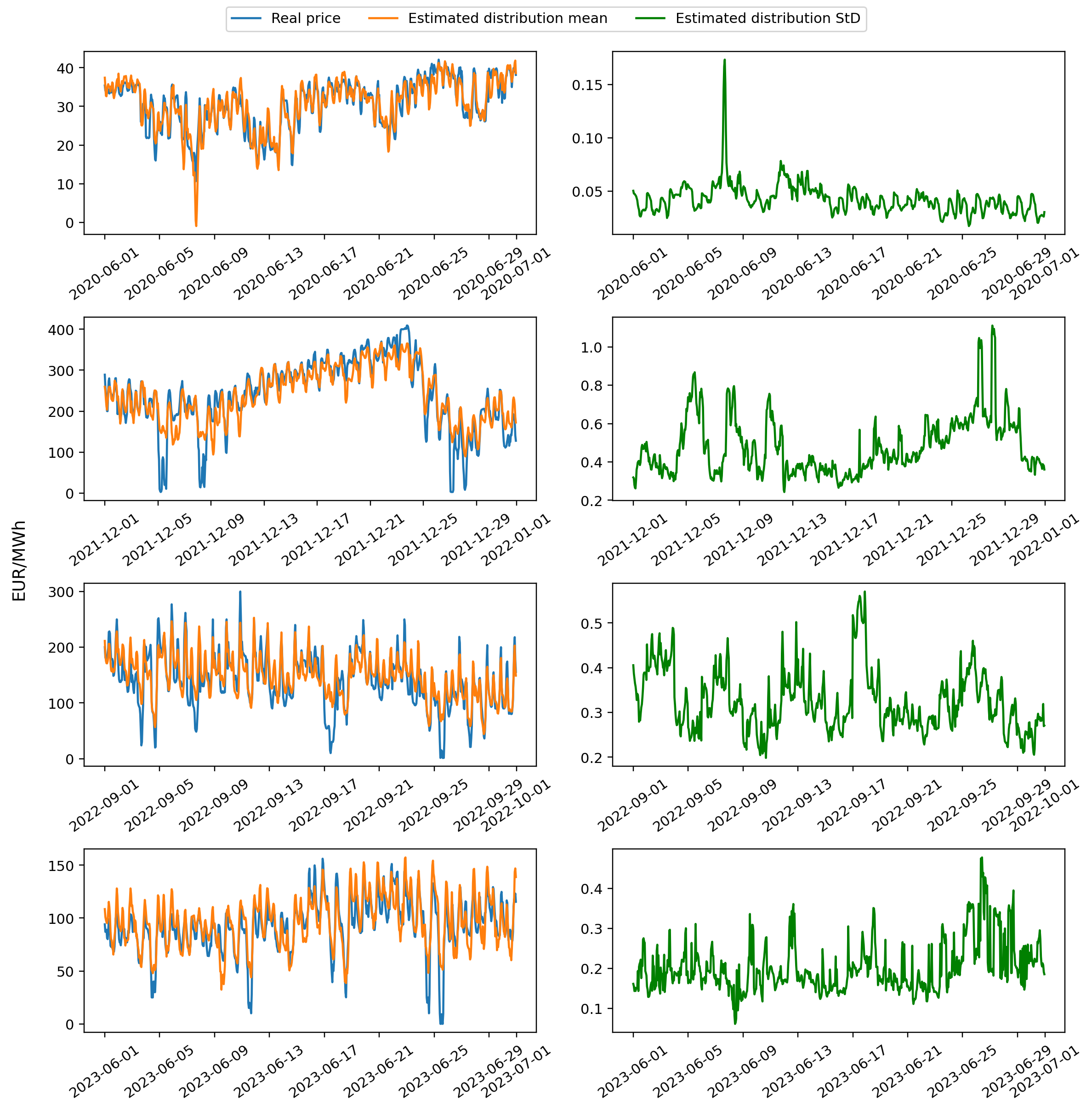}
    \caption{Estimated mean and standard deviation of the hourly price distribution on four different periods (June 2020, December 2021, September 2022, and June 2023).}
    \label{fig:est_mean_sd}
\end{figure}

Now, we check whether the estimated mean from the distribution is far from the real price, and examine the value of the estimated standard deviation, which is directly related to the uncertainty around the prediction. We can check these estimated distribution parameters in Figure \ref{fig:est_mean_sd}.

Here, we present the estimated distribution mean and standard deviations on four different periods: June 2020 (low prices during the COVID period), December 2021 (high market prices due to oil and gas prices), September 2022 (still high prices with inflation increasing), and June 2023 (more stable period, but prices around double than before COVID).

If we focus on the estimated mean, we obtain a mean absolute error (MAE) of 1.8\euro/MWh on June 2020, 25.8\euro/MWh on December 2021, 20.1\euro/MWh on September 2022, and 11.2\euro/MWh on June 2023. As we expected, extreme periods are harder to forecast, and therefore we obtain a higher MAE, such as in December 2021. However, we can notice that, when the estimated distribution mean lies far from the real price, the estimated distribution standard deviation increases. This means that the DNN detects that the specific hourly price distribution will be hard to estimate, and therefore the standard deviation value is increased. This is an insightful result, as we know the GENCO will face high levels of uncertainty at the time the standard deviation of the distribution is higher.

With these results, we assert that the DNN is capable of modeling the behavior of the market given some contextual information and the block information of the GENCO. In this sense, we assume that GENCO decisions can affect the marginal price, and hence, their profits (which will be tested next).

\subsubsection{Scenario generation}

After embedding the trained DNN within our optimization problem, the last step is to generate scenarios $\omega \in \Omega$. As it has been stated in Section \ref{sec:methodology}, the scenario uncertainty will come from the fulfillment of constraint (\ref{eq:dcl_yrandom}), employing randomly generated values from a standard normal distribution. A total of 150 of these sampled values will be used, which implies a total of 150 scenarios per hour in the test month.

\subsection{Results}

We solve the stochastic optimization problem (\ref{eq:stochastic_model_full2}) through the scenario generation described in the previous section. This optimization problem will be solved for each day within the month of June 2023. In the Spanish market, producers send their offers at 12:00 for every hour of the following day. For that reason, the quantity offered per block ($Q_t^{\text{Max }i}$ ) and its true estimated cost ($C_t^i$) will be determined by solving the model (\ref{eq:discr_model}) to obtain the respective blocks for the GENCO.

The proposed optimal offering problem has been solved through a Python 3.9.12 implementation, using Pyomo 6.3 \parencite{hart2017pyomo}. The selected mathematical solver for all the computations was Gurobi \parencite{gurobi} in its version 10.0. Besides, the computer employed included a CPU Intel Core i7 10700, RAM of 64 GB, and NVIDIA GeForce GTX 3060 graphic card. The computation time is dependent on the particular scenario set and the assigned parameters, especially $\sigma_t^i$, where a bigger value increases the decision space, and therefore, the computation time. In general, most computationally demanding experiments took around 1 hour per day of the testing set to reach a 0.5\% optimality gap.

As indicated, we use $\sigma_t^i$ to quantify possible deviations with respect to $C_t^i$. For example, if the production cost of the first block at time $t$ is 25\euro, and we set a variability of 10\%, $\sigma_t^i$ will have a value of 2.5\euro. From now on, we will directly refer to $\sigma_t^i$ as this percentage. This variability has been applied to the first four non-zero cost blocks. For the remaining two blocks, $\sigma_t^i$ was set to zero, as their cost is so high that their price never intervenes in the marginal price formation. Moreover, the $\alpha$ value affecting the CVaR formulation is set to 10\%, that is, in the risk-averse case, we focus on improving the expected value of the scenarios below the 10th percentile of the profit distribution.

Regarding the rest of the covariates needed by the DNN to estimate the marginal price, they are also known at the time of the decision-making, as the  ESIOS platform offers open-access estimations of hourly demand, renewable production, etc. Thus, we consider this problem as a realistic data-driven approach.

Concerning the risk aversion level, we solve the optimization problem for the cases where the producer is risk-neutral ($\chi = 1$) and risk-averse ($\chi = 0$). In the computation process, we used $\chi$ values 0.001 and 0.999 instead of 0 and 1 for stability in the results.

\subsubsection{Aggregated results}

First, we present a summary of the stochastic results for $\chi$ values (risk aversion level) of zero and one, and $\sigma_t^i$ values of $5\%$, $10\%$ and $15\%$, in Table \ref{tab:stoch_results}. Notice that the employment of $0\%$ as the value of $\sigma_t^i$ leads to a heuristic approach where the GENCO does not modify her offering prices in comparison to her costs.

The idea of limiting $\sigma_t^i$ to a specific set of small values comes from the employment of the DCL methodology. Allowing the GENCO to deviate too much from her cost can cause optimal solutions far from the data that was previously employed to train the predictive model, which may cause a risk of increasing the modeling error. Furthermore, we believe that in a complex, large, and marginal energy market, regulators will try to supervise large producers' offering behavior, not allowing large deviations from their true costs (large value of $\sigma_t^i$). In the same sense, we believe that the aforementioned deviation values are enough to study GENCO's optimal offering, which is later validated out-of-sample.

Furthermore, we introduce two additional approaches for comparative purposes. First, what we denote as Deterministic Constraint Learning (Det. CL) transforms the stochastic problem into a deterministic one. In this sense, the NN will only output the mean estimate of the conditional distribution for the marginal price. This point prediction will be used to solve the problem (\ref{eq:stochastic_model_full}) in a deterministic way, as no scenarios are generated. Once the optimal deterministic offering curve solution is obtained, it will be fixed and evaluated as a stochastic problem.

On the other hand, we will also compare our DCL methodology with its ``wait and see'' (perfect information) counterpart. In this case, the GENCO can anticipate the exact realization of the marginal price, which directly depends on the random values $z_{\omega}$ in the constraint (\ref{eq:dcl_yrandom}). A different optimal offering strategy will be obtained for each of the scenarios, and therefore reach the maximum possible profit assuming a perfect foresight of the future. This approach cannot be implemented in practice, but can be used as a benchmark to evaluate the performance of the stochastic model.

The three first columns of Table \ref{tab:stoch_results} represent the considered approach, the price flexibility over the cost of production, and the risk aversion level of the GENCO, respectively. The expected profit is calculated as the mean daily profit over the 30 days of testing; the same testing period is used to derive the expected CVaR. The sixth column represents the expected learned marginal price during June 2023, while the seventh column reports the expected dispatched energy, that is, the energy to be produced as its offered price is under the market marginal price. Finally, the last four columns present the mean price offered for each of the first four blocks (decision variables). Notice that, for the perfect information approach, we obtain a different offering curve per scenario.

\begin{table}[ht]
\caption{Summary of results for the GENCO's optimization problem.}
\centering
\label{tab:stoch_results}
\resizebox{\textwidth}{!}{%
\begin{tabular}{ccc|cccccccc}
\multicolumn{1}{p{1.5cm}}{\centering Solving \\ Approach}  & \multicolumn{1}{p{1.5cm}}{\centering Price \\ flexibility}  &  \multicolumn{1}{p{1.5cm}|}{\centering Risk \\ aversion}     & \multicolumn{1}{p{2cm}}{\centering $\mathbb{E}[\text{Profit}]$ \\ (\euro)}    & \multicolumn{1}{p{2cm}}{\centering $\mathbb{E}[\text{CVaR}]$  \\ (\euro)} & \multicolumn{1}{p{1.5cm}}{\centering $\mathbb{E}[\lambda_{t,\omega}]$  \\ (\euro/MWh) } & \multicolumn{1}{p{2.75cm}}{\centering $\mathbb{E}[Q^{ren}_t + \sum Q_{t,\omega}^{i}]$  \\ (MWh) } & \multicolumn{1}{p{1.5cm}}{\centering $\mathbb{E}[P_t^1]$ \\ (\euro/MWh) } & \multicolumn{1}{p{1.5cm}}{\centering $\mathbb{E}[P_t^2]$ \\ (\euro/MWh) }   & \multicolumn{1}{p{1.5cm}}{\centering $\mathbb{E}[P_t^3]$ \\ (\euro/MWh) }    & \multicolumn{1}{p{1.5cm}}{\centering $\mathbb{E}[P_t^4]$ \\ (\euro/MWh) }   \\ \hline

Cost-based & \multirow{1}{*}{$\sigma_t^i = 0\%$}      & $\forall \chi$ & 4,589,456 & 3,270,420 & 96.74 & 2,506 & 60.85 & 81.72 & 96.91 & 136.08 \\ \hline \hline

\multirow{3}{*}{Det. CL} &   \multirow{1}{*}{$\sigma_t^i = 5\%$}      & $\forall \chi$ & 4,560,588 & 3,236,255 & 96.32 & 2,542 & 57.87 & 81.03 & 93.11 & 129.56 \\  \cline{2-11}
& \multirow{1}{*}{$\sigma_t^i = 10\%$}     & $\forall \chi$ & 4,538,147 & 3,203,301 & 96.14 & 2,562 & 55.08 & 81.40 & 92.16 & 124.06 \\ \cline{2-11}
& \multirow{1}{*}{$\sigma_t^i = 15\%$}     & $\forall \chi$ & 4,513,302 & 3,164,629 & 95.99 & 2,555 & 52.18 & 81.76 & 93.30 & 121.52 \\  \hline \hline

\multirow{6}{*}{DCL} &   \multirow{2}{*}{$\sigma_t^i = 5\%$}  & $\chi = 0$ & 4,644,115 & 3,317,245 & 97.68 & 2,482 & 63.46 & 85.78 & 101.28 & 135.16 \\
                                     &    & $\chi = 1$ & 4,638,018 & 3,324,961 & 97.59 & 2,485 & 63.38 & 85.41 & 101.02 & 136.44 \\ \cline{2-11}
& \multirow{2}{*}{$\sigma_t^i = 10\%$}    & $\chi = 0$ & 4,693,195 & 3,354,528 & 98.58 & 2,457 & 65.94 & 89.79 & 104.95 & 135.98 \\
                                    &     & $\chi = 1$ & 4,679,540 & 3,373,450 & 98.40 & 2,460 & 65.72 & 89.03 & 104.67 & 138.22 \\ \cline{2-11}
& \multirow{2}{*}{$\sigma_t^i = 15\%$}     & $\chi = 0$ & 4,736,825 & 3,384,990 & 99.42 & 2,437 & 68.37 & 93.58 & 107.82 & 137.63 \\
                                  &       & $\chi = 1$ & 4,710,707 & 3,410,710 & 99.03 & 2,441 & 67.52 & 92.05 & 107.53 & 141.14 \\ \hline \hline

\multirow{3}{*}{W\&S DCL} &   \multirow{1}{*}{$\sigma_t^i = 5\%$}      & $\forall \chi$ & 4,692,449 & 3,341,055 & 98.13 & 2,496 & 62.90 & 85.57 & 101.22 & 135.80 \\  \cline{2-11}
& \multirow{1}{*}{$\sigma_t^i = 10\%$}     & $\forall \chi$ & 4,747,639 & 3,391,311 & 99.05 & 2,496 & 65.87 & 89.35 & 105.27 & 136.37 \\ \cline{2-11}
& \multirow{1}{*}{$\sigma_t^i = 15\%$}     & $\forall \chi$ & 4,800,898 & 3,437,531 & 99.95 & 2,495 & 67.71 & 93.07 & 109.07 & 138.18 \\  \hline \hline

\end{tabular}
}
\end{table}

Regarding the deterministic approach, we can see how the expected profit decreases as the price flexibility level increases. This could be due to not taking uncertainty into account, as a point estimation maximization could lead to higher uncertainty (${\lambda_{t}}$) in the stochastic solution. However, more dispatched energy is obtained compared to the DCL approach. With the perfect information scenario, a higher profit is expected, but not far from the uncertain DCL approach. It is interesting to note that the CVaR of the risk-averse GENCO with the DCL approach is quite close to the one obtained in the perfect information case.

By differentiating between levels of price flexibility within the DCL approach, we can see how the expected profit increases as this flexibility does. This is due to the offering price adjustments that the GENCO employs to modify the marginal price of the market. Starting from a marginal price of 96.74\euro/MWh when the GENCO makes offers at production cost, we can see an increment of the marginal price up to 99.42\euro/MWh where the flexibility is 15\% of the production costs.

There are also differences regarding the risk aversion level. When the producer is risk-averse ($\chi = 1$), expected profits are slightly lower, but the CVaR improves significantly, as the marginal price does not increase as much as in the case where the producer is risk-neutral ($\chi = 0$). The dispatched energy differences across flexibility and risk aversion levels are small, making us guess that the GENCO is more interested in modifying the marginal price than increasing its dispatched quantities.

Concerning the block prices offered, significant differences appear when varying the price flexibility and risk aversion levels. For example, as flexibility increases, the mean price offered for the fourth block increases in the risk-averse GENGO, while maintaining its value closer to its cost for the risk-neutral case, which could be due to the risk of getting blocks dispatched for a price lower than their costs. On the other hand, the price of the first, second, and third blocks increase independently of the risk-aversion level with a larger flexibility, but this increment is on a lower scale on the risk-averse GENCO. This fact can be caused by the need for the producer to get some blocks dispatched, and their preference for getting the blocks dispatched against increasing even more the marginal price.

\subsubsection{Results by risk-aversion and flexibility level}

What follows will graphically show the stochastic results for the complete testing period, studying the profit distribution and optimal block prices for different levels of risk aversion and price flexibility.

\begin{figure}[ht]
    \centering
    
    \begin{subfigure}{0.49\textwidth}
    \centering
    \includegraphics[width=\textwidth]{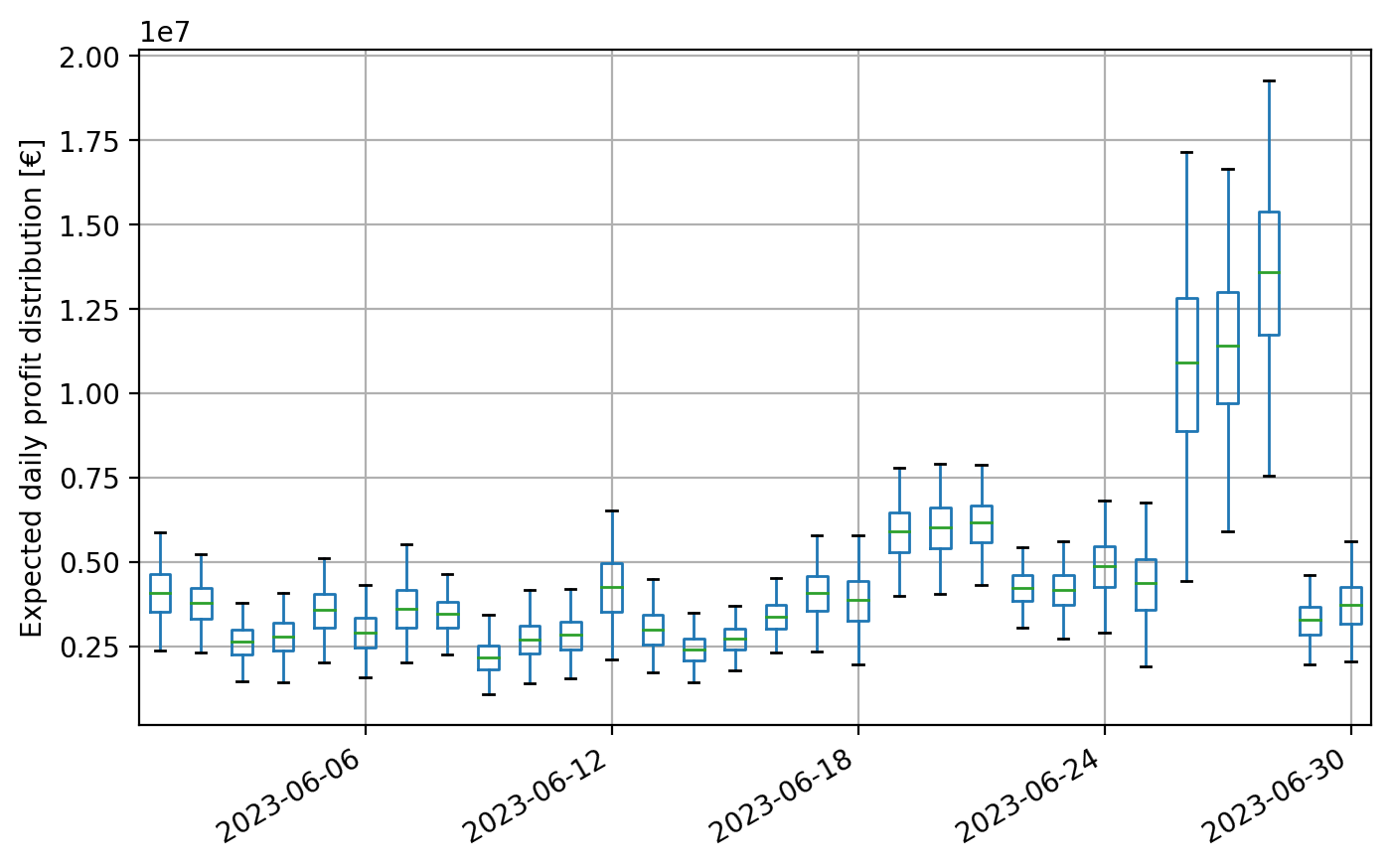}
    \end{subfigure}
    \hfill
    \begin{subfigure}{0.49\textwidth}
    \centering
    \includegraphics[width=\textwidth]{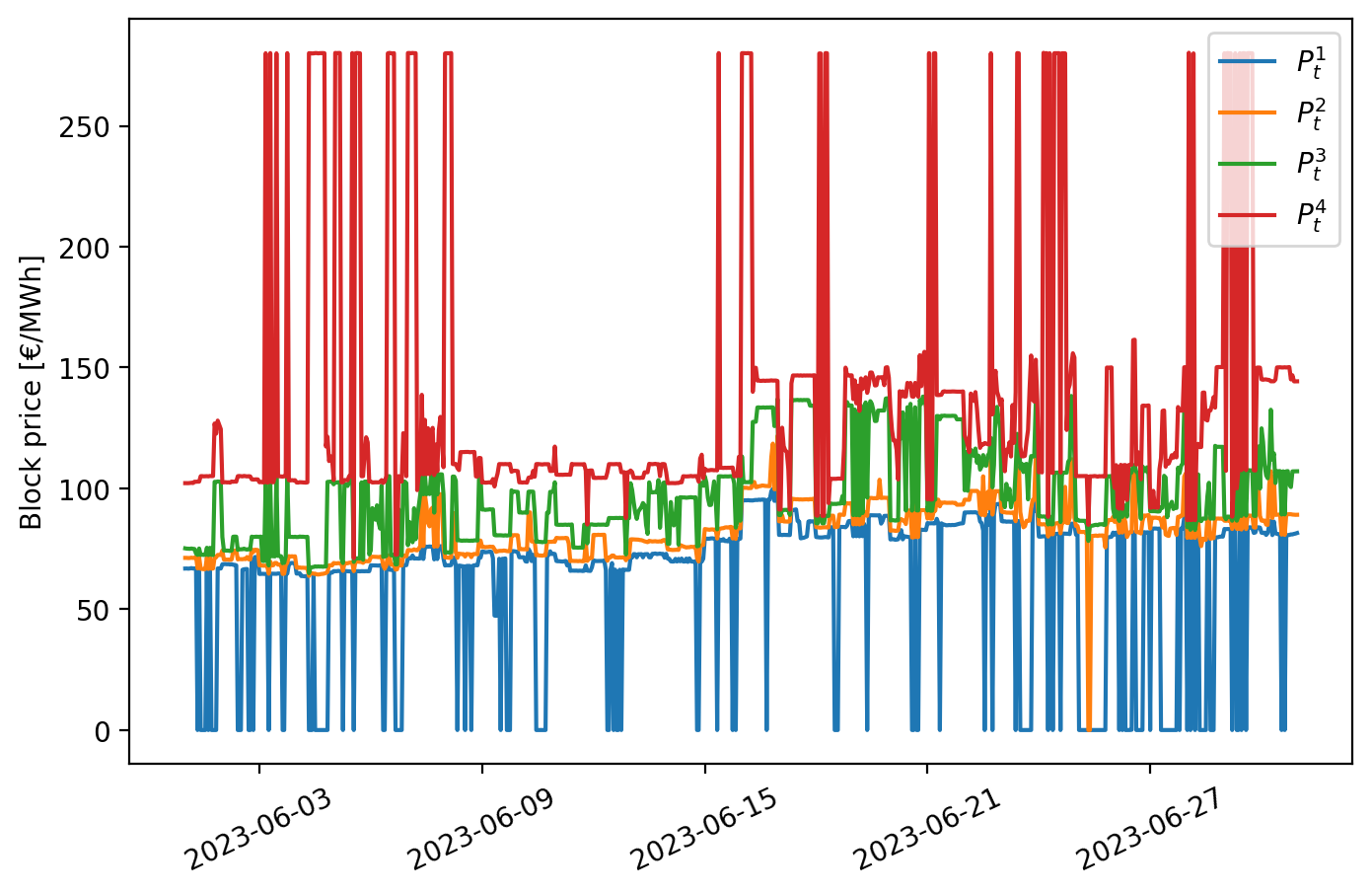}
    \end{subfigure}
    \caption{Expected daily profit distribution (left) and hourly block prices (production costs, right) $\forall \chi$ and $\sigma_t^i = 0\%$}
    \label{fig:risk0_1_sigma_0}
\end{figure}

Firstly, in Figure \ref{fig:risk0_1_sigma_0}, we show the expected daily profit distribution (left) and hourly block prices (right) when no strategic offering is allowed. That is, the price flexibility is 0\% and the block prices are the production costs. With this level of price flexibility, the block prices show the real production cost during June 2023. These costs are obtained following the methodology presented in Section \ref{sec:supply_disc}. That is, we discretized the original supply curve of the large GENCO and assumed the obtained values for each block to be the true costs. Regarding the expected profit distribution, we can see how it fluctuates between 2 and almost 15 million euros on some specific days. Furthermore, we can see the different profit uncertainty levels the GENCO faces across the month. For example, the profit distribution at the beginning of the month is much narrower than at the end, giving us an idea of different levels of risk exposure.

\begin{figure}[ht]
    \centering
    \begin{subfigure}{0.49\textwidth}
    \centering
    \includegraphics[width=\textwidth]{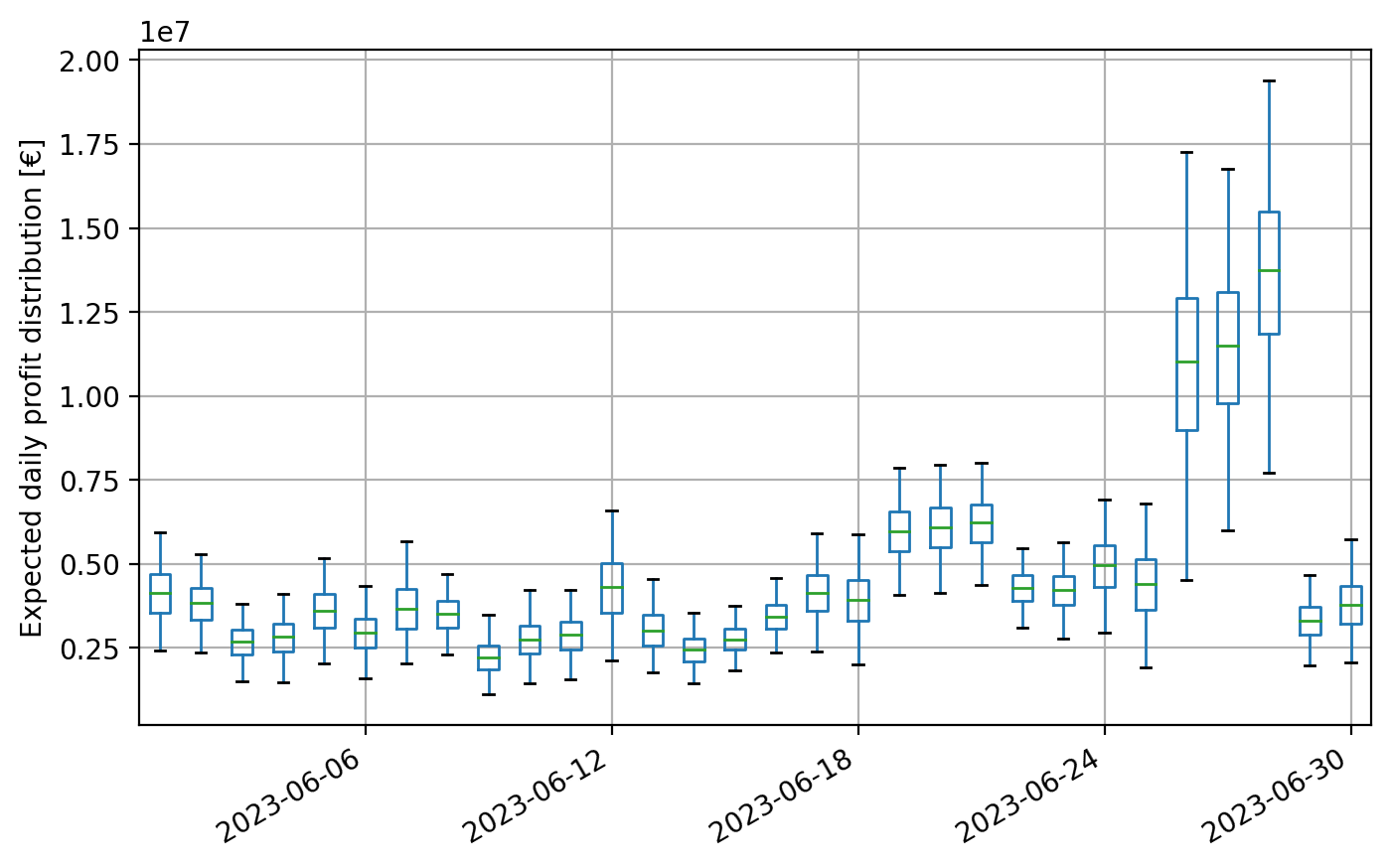}
    \caption{Distribution of daily profit, $\chi=0$, $\sigma_t^i = 5\%$}
    \end{subfigure}
    \hfill
    \begin{subfigure}{0.49\textwidth}
    \centering
    \includegraphics[width=\textwidth]{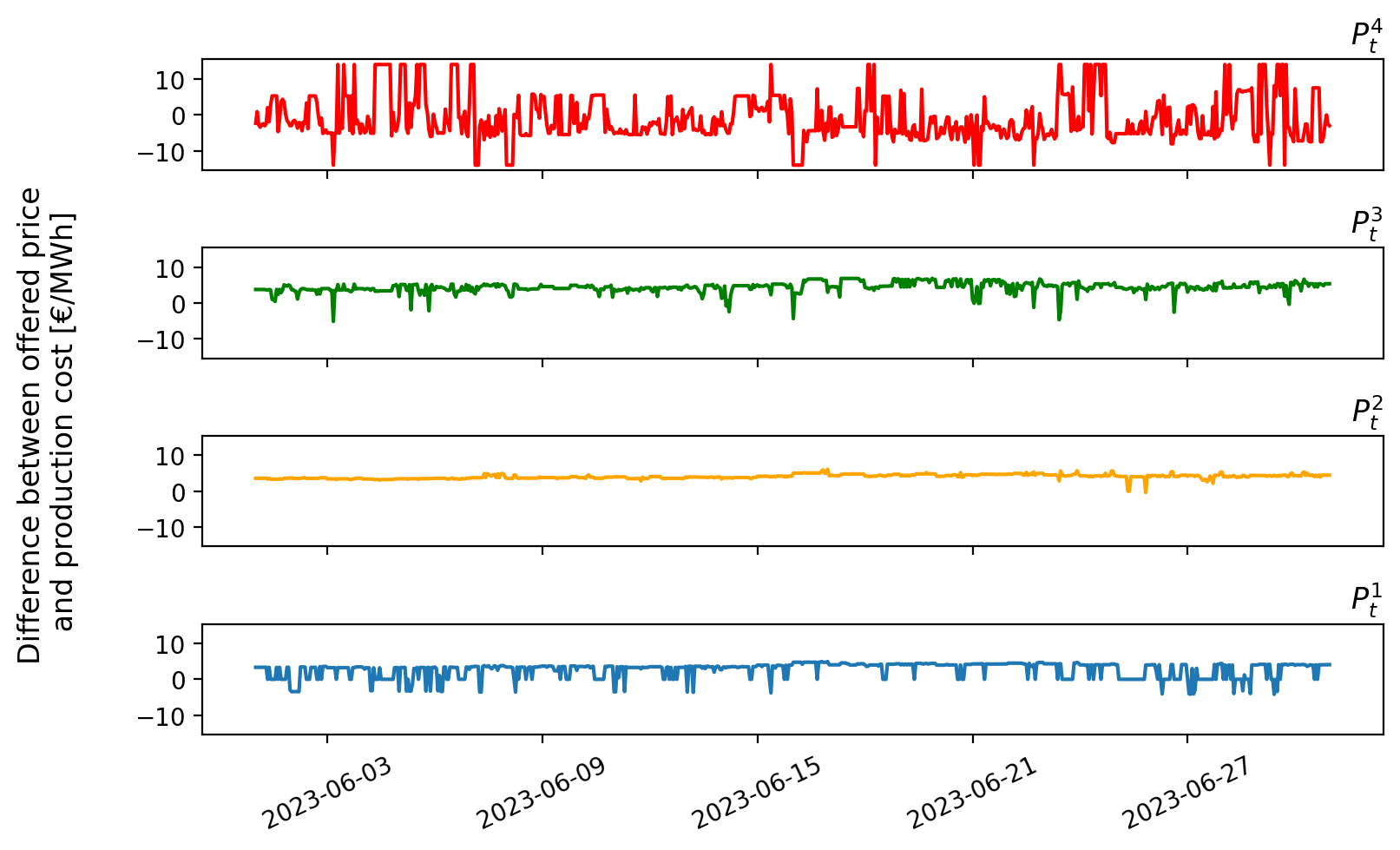}
    \caption{Block prices vs. production costs, $\chi=0$, $\sigma_t^i = 5\%$}
    \end{subfigure}
    \centering
    \begin{subfigure}{0.49\textwidth}
    \centering
    \includegraphics[width=\textwidth]{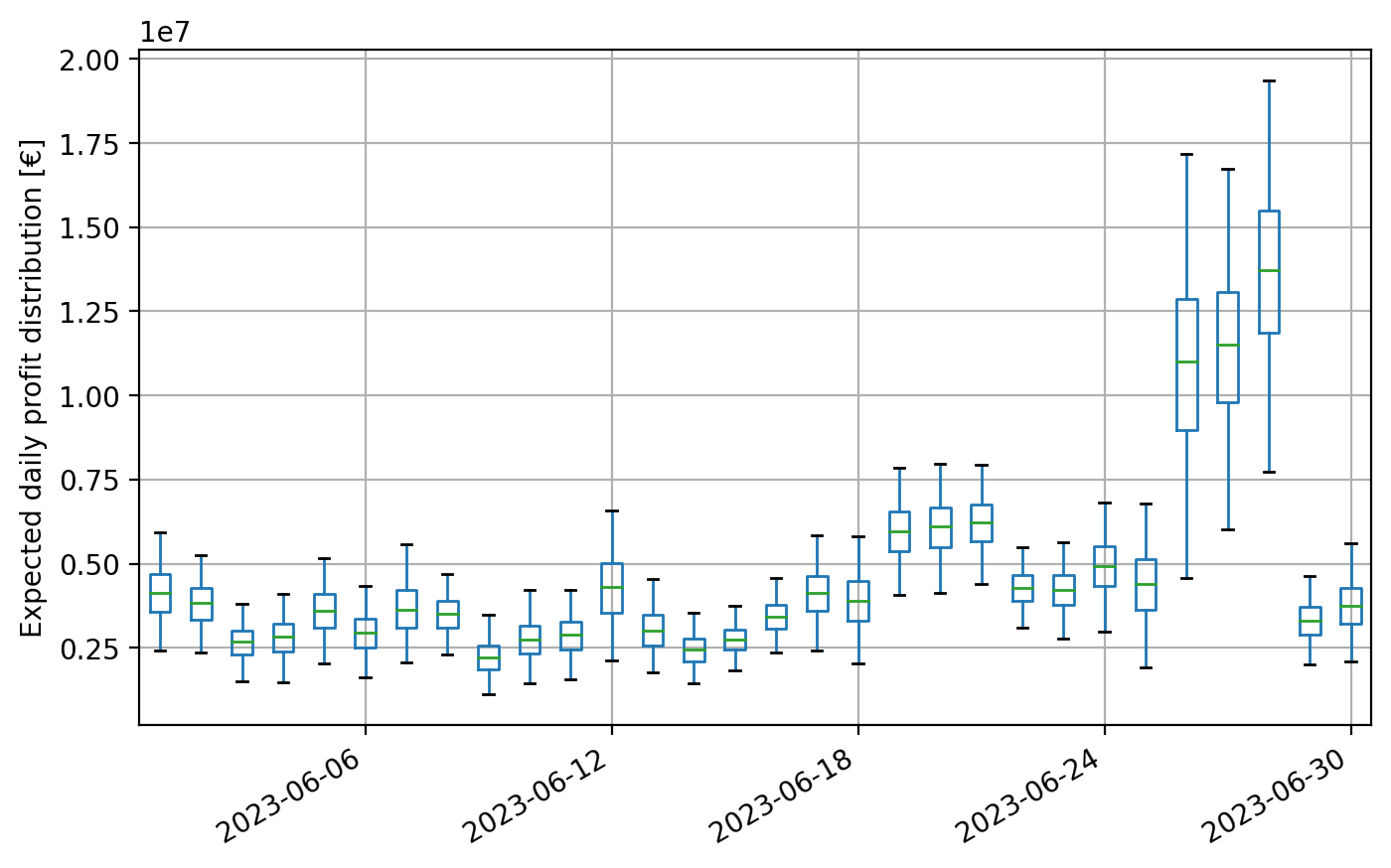}
    \caption{Distribution of daily profit, $\chi=1$, $\sigma_t^i = 5\%$}
    \end{subfigure}
    \hfill
    \begin{subfigure}{0.49\textwidth}
    \centering
    \includegraphics[width=\textwidth]{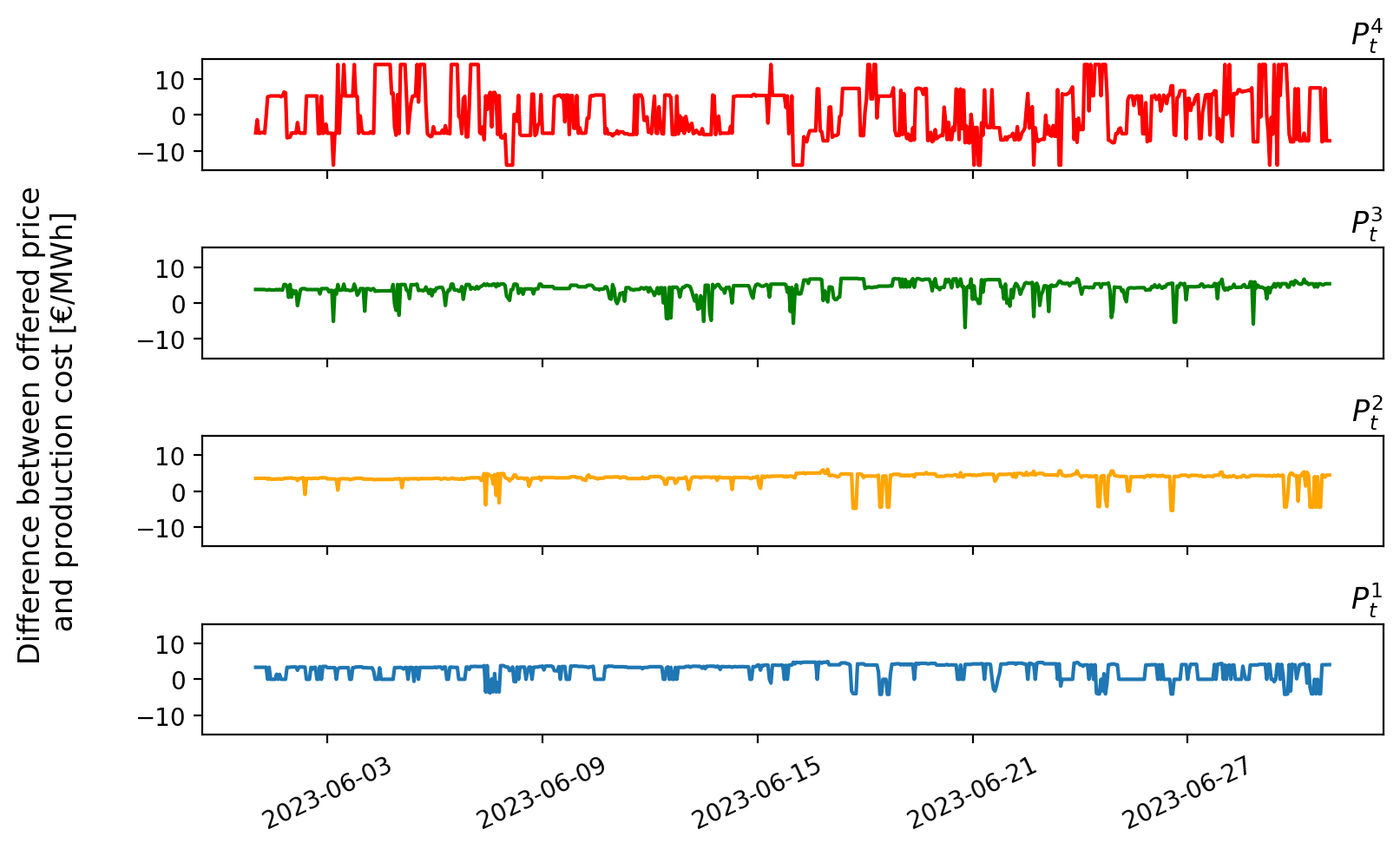}
    \caption{Block prices vs. production costs, $\chi=1$, $\sigma_t^i = 5\%$}
    \end{subfigure}
    \caption{Expected profit distribution and block prices increment for cases $\chi = 0,1$ and $\sigma_t^i = 5\%$}
    \label{fig:risk0_1_sigma_5}
\end{figure}

\begin{figure}[t]
    \centering
    \begin{subfigure}{0.49\textwidth}
    \centering
    \includegraphics[width=\textwidth]{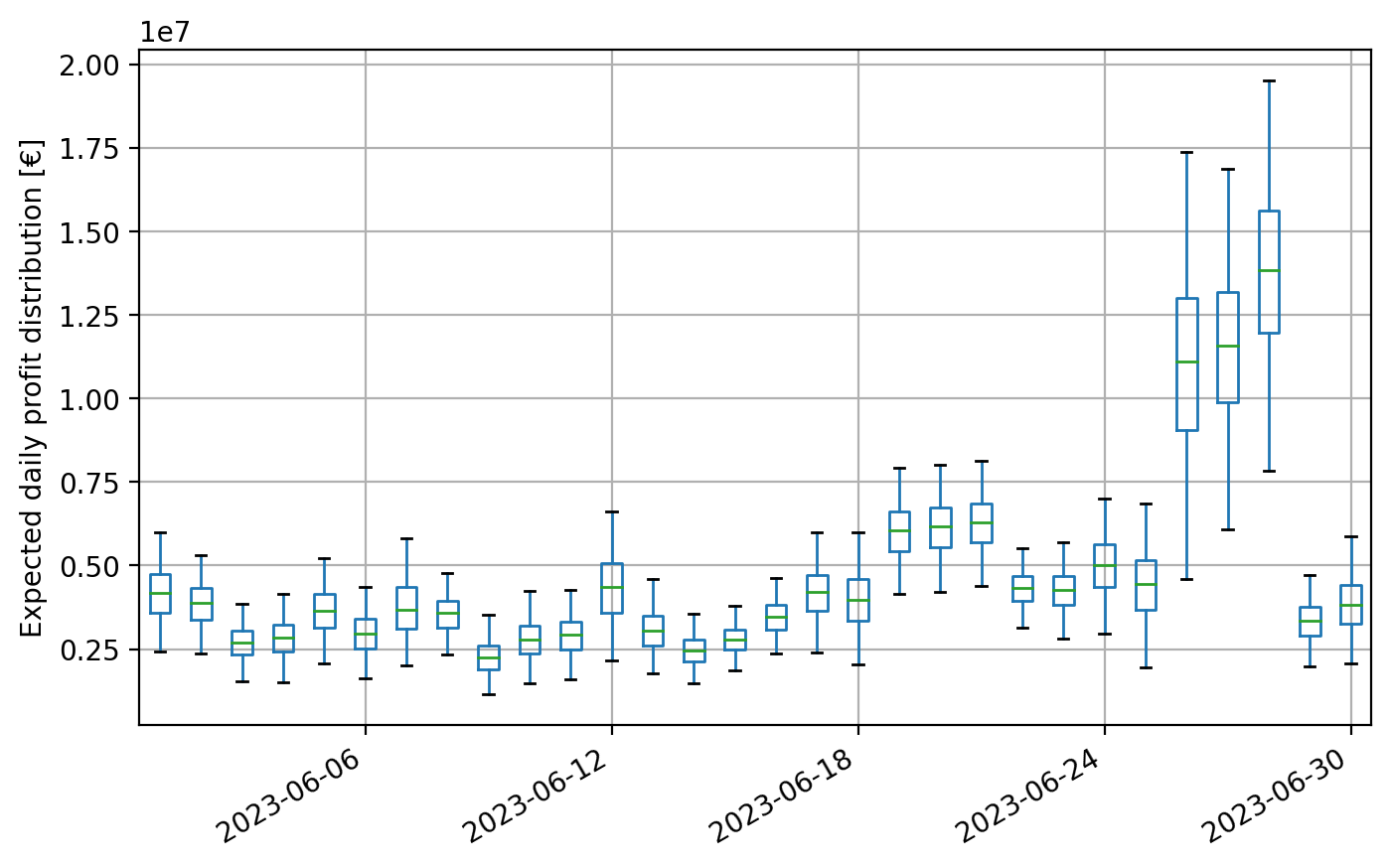}
    \caption{Distribution of daily profit, $\chi=0$, $\sigma_t^i = 10\%$}
    \end{subfigure}
    \hfill
    \begin{subfigure}{0.49\textwidth}
    \centering
    \includegraphics[width=\textwidth]{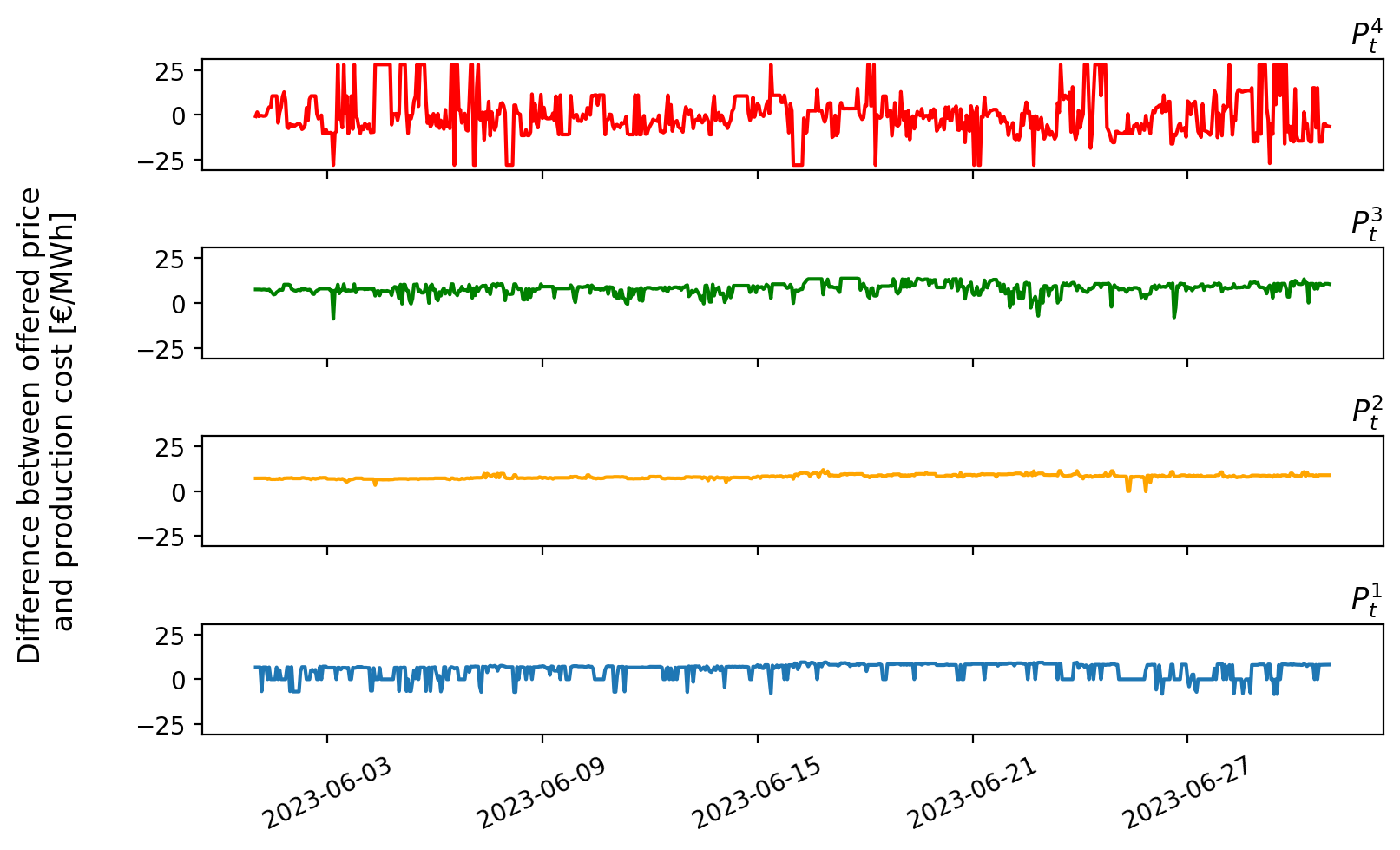}
    \caption{Block prices vs. production costs, $\chi=0$, $\sigma_t^i = 10\%$}
    \end{subfigure}
    \centering
    \begin{subfigure}{0.49\textwidth}
    \centering
    \includegraphics[width=\textwidth]{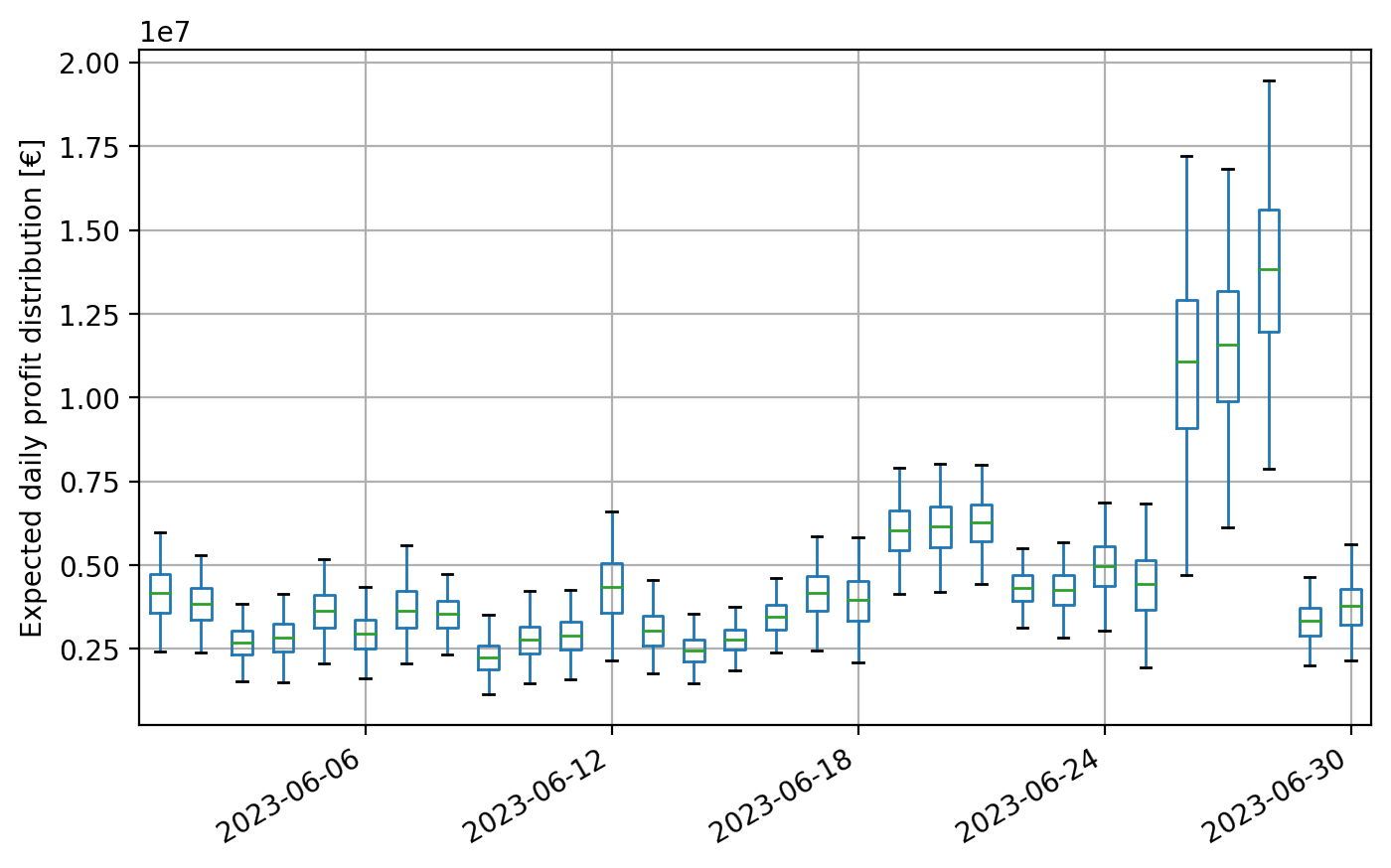}
    \caption{Distribution of daily profit, $\chi=1$, $\sigma_t^i = 10\%$}
    \end{subfigure}
    \hfill
    \begin{subfigure}{0.49\textwidth}
    \centering
    \includegraphics[width=\textwidth]{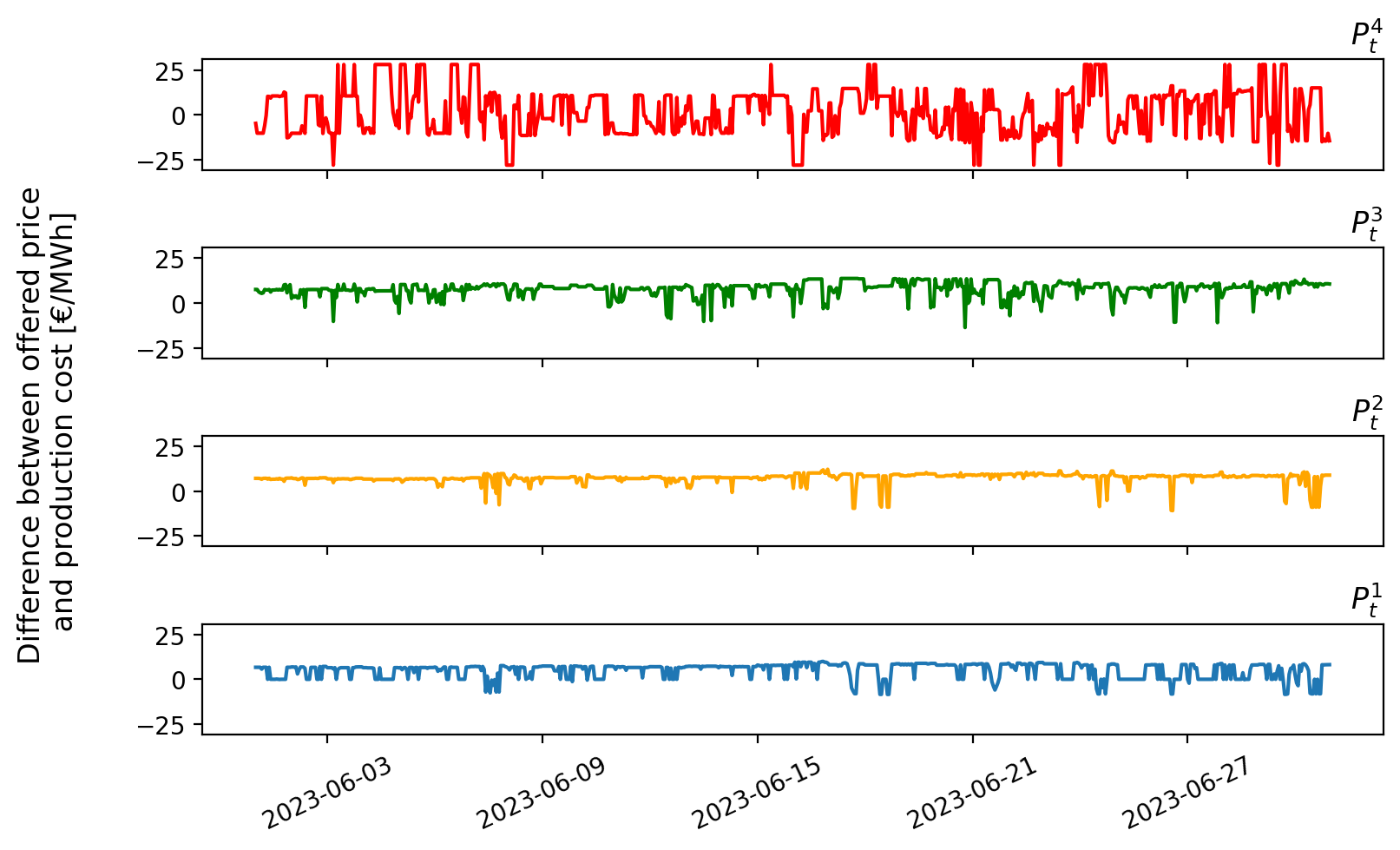}
    \caption{Block prices vs. production costs, $\chi=1$, $\sigma_t^i = 10\%$}
    \end{subfigure}
    \caption{Expected profit distribution and block prices increments for cases $\chi = 0,1$ and $\sigma_t^i = 10\%$}
    \label{fig:risk0_1_sigma_10}
\end{figure}

\begin{figure}[t]
    \centering
    \begin{subfigure}{0.49\textwidth}
    \centering
    \includegraphics[width=\textwidth]{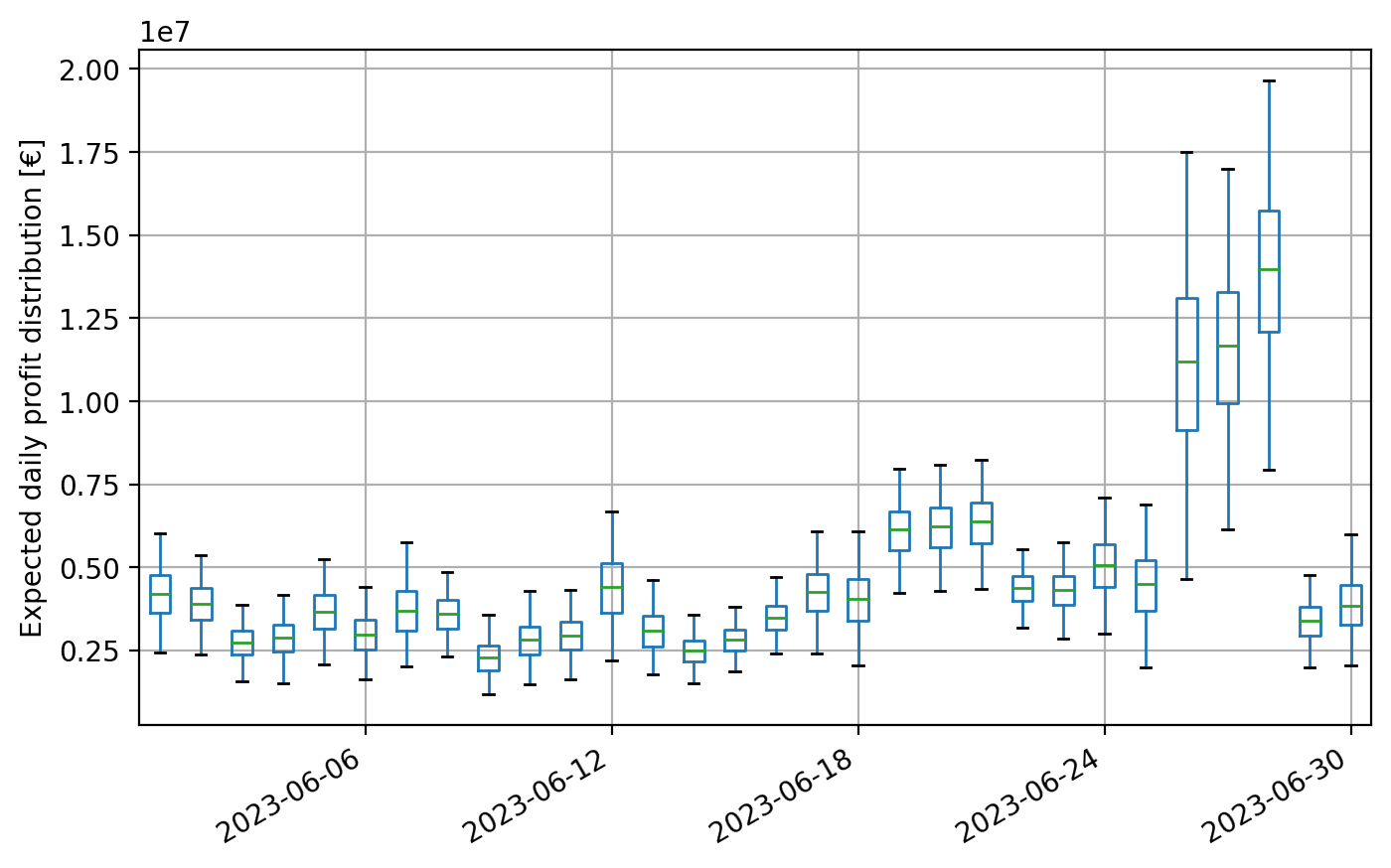}
    \caption{Distribution of daily profit, $\chi=0$, $\sigma_t^i = 15\%$}
    \end{subfigure}
    \hfill
    \begin{subfigure}{0.49\textwidth}
    \centering
    \includegraphics[width=\textwidth]{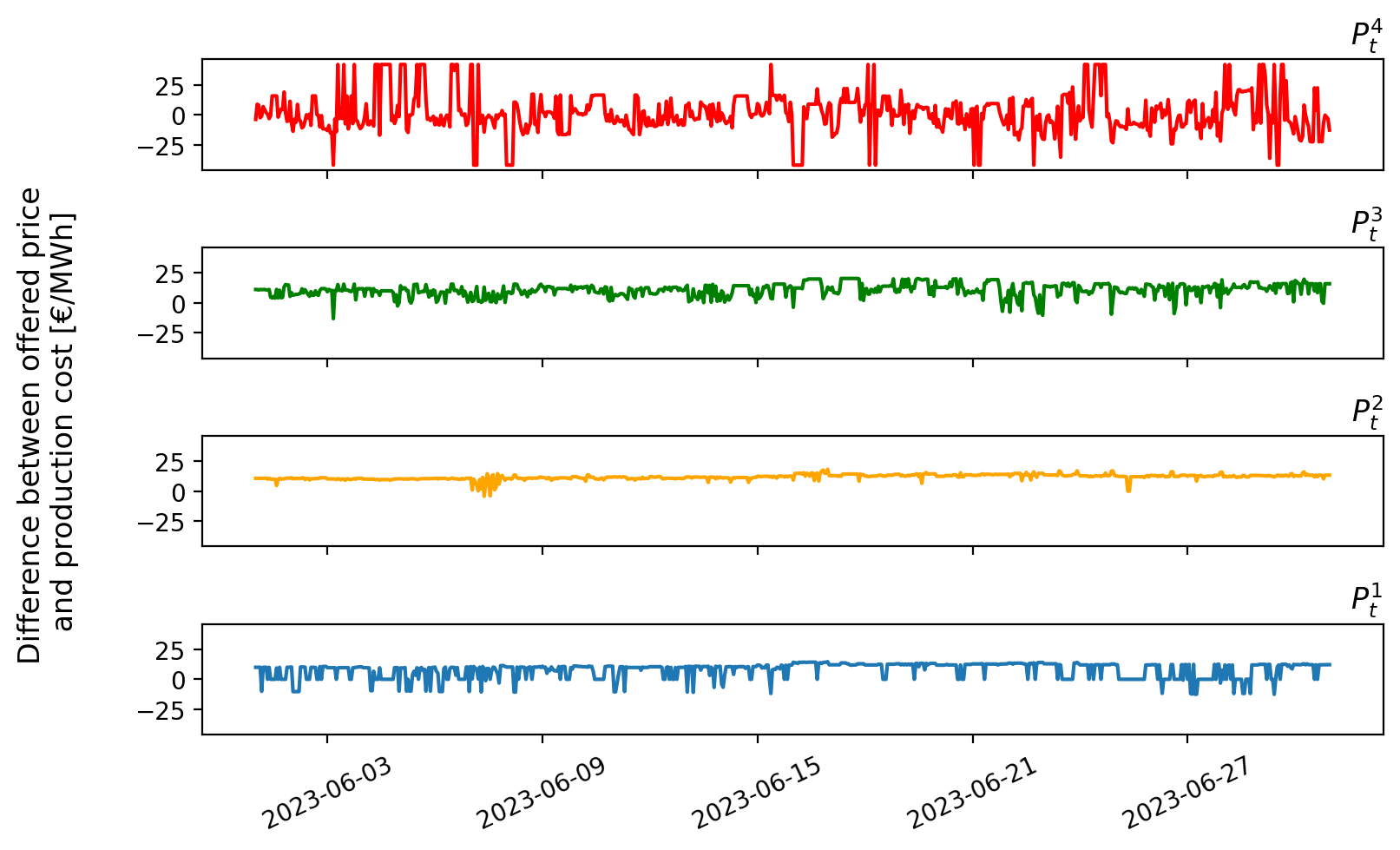}
    \caption{Block prices vs. production costs, $\chi=0$, $\sigma_t^i = 15\%$}
    \end{subfigure}
    \centering
    \begin{subfigure}{0.49\textwidth}
    \centering
    \includegraphics[width=\textwidth]{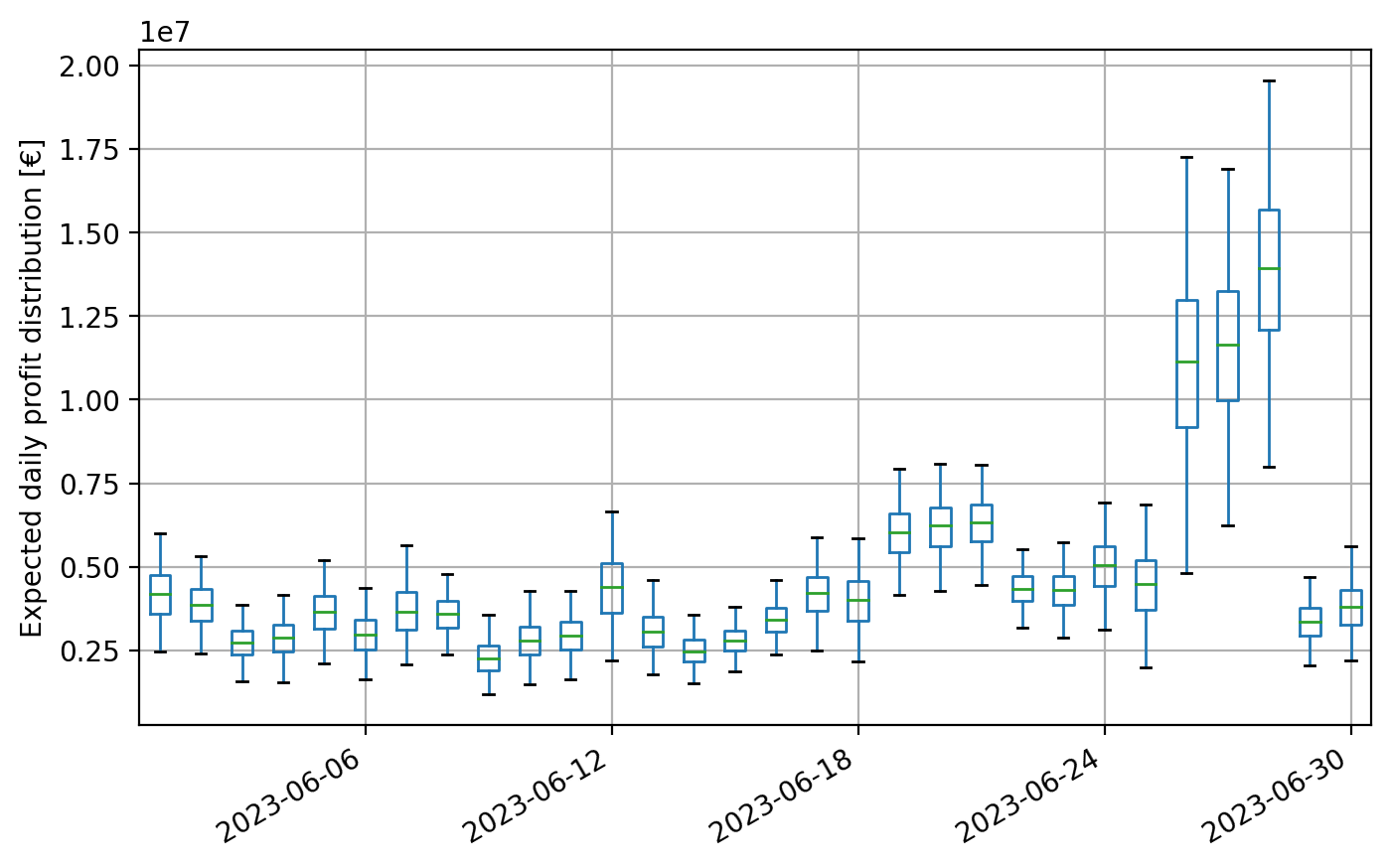}
    \caption{Distribution of daily profit, $\chi=1$, $\sigma_t^i = 15\%$}
    \end{subfigure}
    \hfill
    \begin{subfigure}{0.49\textwidth}
    \centering
    \includegraphics[width=\textwidth]{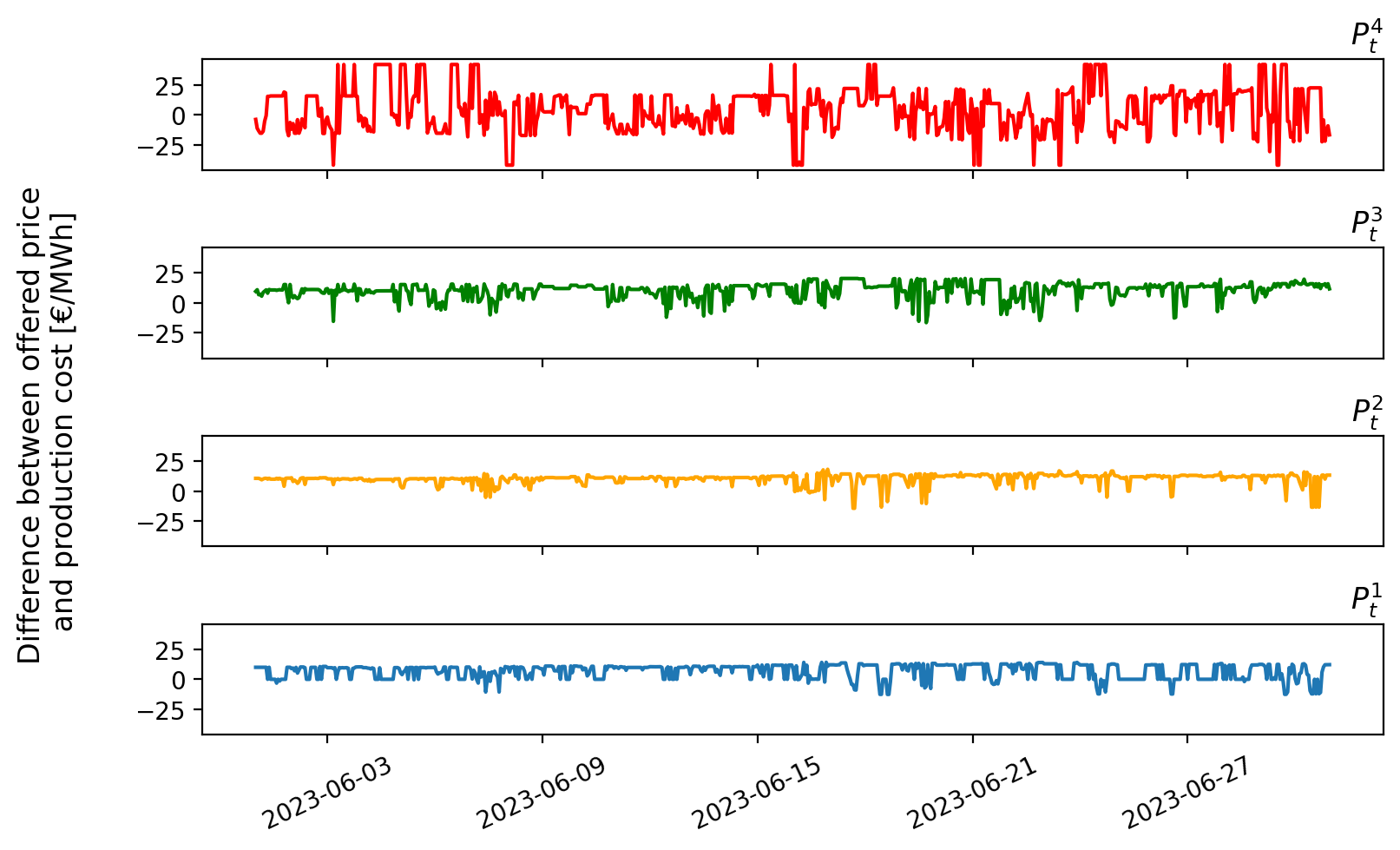}
    \caption{Block prices vs. production costs, $\chi=1$, $\sigma_t^i = 15\%$}
    \end{subfigure}
    \caption{Expected profit distribution and block prices increments for cases $\chi = 0,1$ and $\sigma_t^i = 15\%$}
    \label{fig:risk0_1_sigma_15}
\end{figure}

Next, we increase the price flexibility level up to 5\% and recast the results by risk aversion level in Figure \ref{fig:risk0_1_sigma_5}. We represent the risk-neutral GENCO in Figures \ref{fig:risk0_1_sigma_5}(a) and (b), and the risk-averse in Figures \ref{fig:risk0_1_sigma_5}(c) and (d). Figures \ref{fig:risk0_1_sigma_5}(a) and (c) show the distribution of daily profit increments compared to the base case when the GENCO offers her energy at production costs. This means that, for the same scenario, we compute the difference between the expected profit at a price flexibility level of 5\%, and a level of 0\%. On the other hand, Figures \ref{fig:risk0_1_sigma_5}(b) and (d) represent the difference between the optimal prices offered for each block for a price flexibility level of 5\% and the base case (the production cost in Figure \ref{fig:risk0_1_sigma_0}). The same descriptive setting is applied for the rest of the analyzed cases with different flexibility levels in Figures \ref{fig:risk0_1_sigma_10} and \ref{fig:risk0_1_sigma_15}.

In Figure \ref{fig:risk0_1_sigma_5} we can see differences regarding the profit distribution by risk aversion level. Although there are not so many differences in the mean of the distribution (this mean is slightly bigger for the risk-neutral GENCO), we can appreciate changes in the strategy to obtain these profits. For instance, we can see how the risk-neutral GENCO keeps the price of the second and third blocks higher than their cost but without modifying their value too much. However, the risk-averse GENCO modifies to a larger extent her offers for trying to improve worst-case scenarios. Moreover, the behavior in the fourth block remains unclear; however, the risk-averse GENCO shows a more pronounced variability in its value.

Following the increase of flexibility, we repeat our analysis for the case of $\sigma_t^i = 10\%$ in Figure \ref{fig:risk0_1_sigma_10}. In this case, we can observe the same behavior regarding the distributions of daily profit increment. In relation to the offered prices, both GENCOs take advantage of the price flexibility level by modifying their offers in a larger range regarding the production costs. The pricing behavior for the second and third blocks is still similar to that of the previous case, while in the first block, risk-neutral GENCO shows more variability at the beginning of the month, in contrast to the risk-averse case, which exhibits a higher variability at the end on the month. The fourth block is still the one with the highest variability in its strategy.

We will finish this analysis with a 15\% level of price flexibility case. Results are shown in Figure \ref{fig:risk0_1_sigma_15}. At this flexibility level, we appreciated the largest differences in profit in Table \ref{tab:stoch_results}. However, due to the extent of that increment with respect to the amount of base profit, it is not easy to identify it in the plot. In relation to the block prices, the risk-averse GENCO tries to minimize the price of the third block on more occasions than the risk-neutral while also increasing the price of the second block, which could suggest the target of joining the prices of both blocks. In general, the risk-averse GENCO modifies on more occasions the offering prices, building a strategy with more variance, but with the aim of minimizing worst-case scenarios.

\begin{figure}
    \centering
    \includegraphics[width=0.6\textwidth]{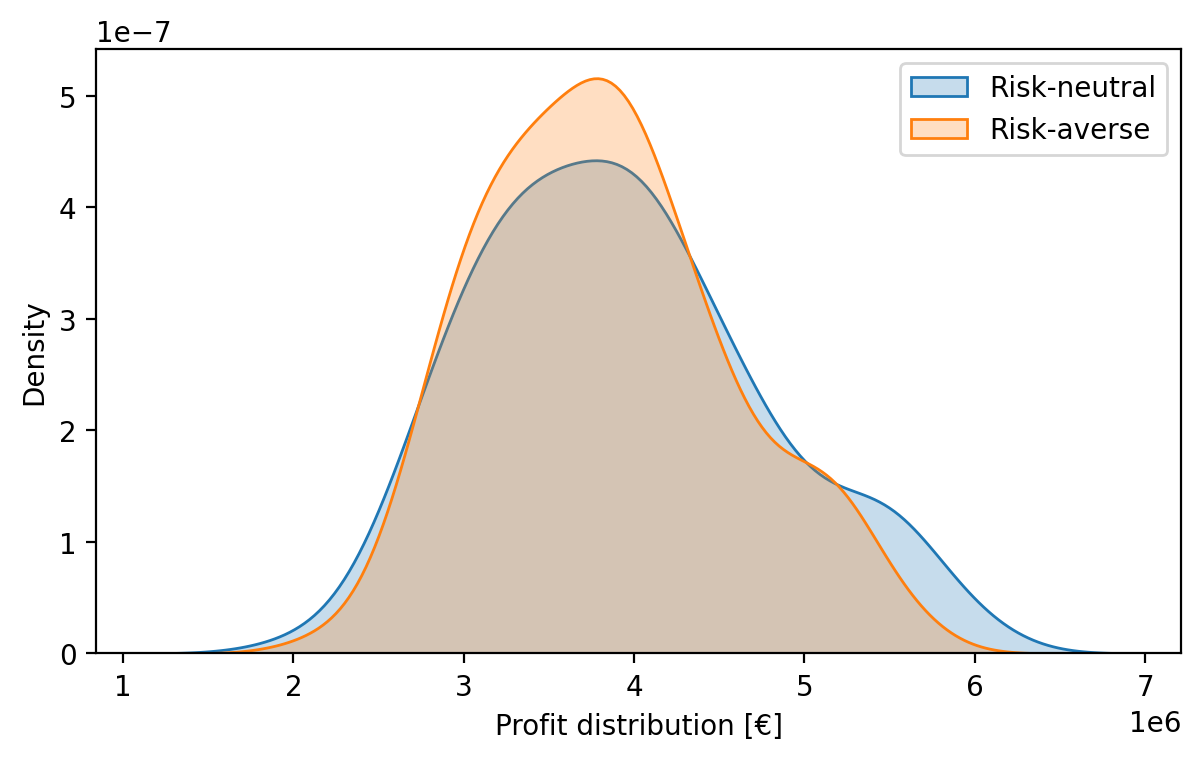}
    \caption{Profit distributions for the risk-neural and risk-averse case (June 30).}
    \label{fig:dif_dist}
\end{figure}

Finally, to better appreciate the difference between profit distributions, we set June 30 as an illustrative example in Figure \ref{fig:dif_dist}. Here, it is easier to check how the risk-averse GENCO aims to improve the worst-case profit scenarios at the cost of decreasing the upper-tail distribution values.

\subsubsection{Out-of-sample model validation}
\label{sec:res_out_of_samp}

As we have seen, the stochastic optimization model illustrates how allowing strategic price offering for a GENCO can increase the day-ahead market marginal price. Furthermore, we see differences in offering prices, marginal prices, and expected profits in relation to the risk aversion level of the GENCO.

This analysis was based on the assumption that the distributional neural network model (\ref{eq:sto_cons_model}) was an accurate representation of the market price response to the GENCO' strategic offers. However, we aim to test if this stochastic optimal strategy would actually work in a real market. That is, we want to know what may occur when the GENCO sends her optimized supply curves to the Spanish day-ahead market. For this reason, we will follow an out-of-sample validation of the derived offering strategy, using data from June 2023.

For this purpose, the methodology will stand as follows:

\begin{enumerate}
    \item Once the optimal strategy by the GENCO is derived from (\ref{eq:stochastic_model_full2}), send the resulting supply curve to the market. Block quantities will be $Q_t^{\text{Max }i}$ with prices $P_t^i$, jointly with $Q_t^{ren}$ at zero cost.
    \item Aggregate the GENCO supply curve with the one collected by the market operator from the rest of the market competitors (also discretizated in 7 blocks, see Section \ref{sec:methodology}), and use the inelastic demand forecast from ESIOS to obtain the initial marginal price.
    \item Displace the market supply curve employing the mean displacement of the last two months at its corresponding hour, to reproduce technical adjustments performed by the market operator.
\end{enumerate}

After all these steps are done, we will finally get an estimated marginal price. Profits for the GENCO can be computed as the product of the dispatched quantities (those blocks below the resulting marginal price) times the marginal price minus the production costs of this energy.

These out-of-sample results have been computed for different levels of risk aversion ($\chi$) and price flexibility ($\sigma_t^i$). Table \ref{tab:oos_results} summarizes the main obtained insights. The third column of the table shows the total daily profit increments for the GENCO in June 2023 when deviating her optimal offering from the production costs. The fourth and fifth columns indicate the first and third quantiles over this daily profit increment distribution. These two values will give us an interval guess of what value we should expect the profit increment to be. We find its standard deviation rightwards, to finally compute the mean hourly marginal price of the market.

\begin{table}[ht]
\caption{Result summary of the out-of-sample model validation.}
\centering
\label{tab:oos_results}
\resizebox{\textwidth}{!}{%
\begin{tabular}{cc|ccccc}
\multicolumn{1}{p{1.5cm}}{\centering Price \\ flexibility}  &  \multicolumn{1}{p{1.5cm}|}{\centering Risk \\ aversion}     & \multicolumn{1}{p{2cm}}{\centering $\sum \Delta\text{Profit}$ \\ ($\text{\euro}$)}  & \multicolumn{1}{p{2cm}}{\centering $Q_1$ \\ ($\text{\euro}$)} & \multicolumn{1}{p{2cm}}{\centering $Q_3$ \\ ($\text{\euro}$)} & \multicolumn{1}{p{2cm}}{\centering $Std[ \Delta \text{Profit}]$ \\ ($\text{\euro}$)}  & \multicolumn{1}{p{2cm}}{\centering $\overline{\lambda_t}$ \\ ($\text{\euro}$/MWh)} \\ \hline

\multirow{2}{*}{$\sigma_t^i = 5\%$}        & $\chi=0$ & 424,103.36  & -507.24 & 20,237.61  & 25,210.43  & 96.91 \\
                                            & $\chi=1$ & 311,326.79 & -2,299.35 & 15,082.07 & 23,862.25 & 96.85   \\ \hline
\multirow{2}{*}{$\sigma_t^i = 10\%$}       & $\chi=0$ & 825,931.25 & -2,085.10  & 41,687.05 & 48,014.47 & 97.19  \\
                                            & $\chi=1$ & 616,259.22 & -4,315.41  & 33,792.90 & 49,863.25 & 97.08 \\ \hline
\multirow{2}{*}{$\sigma_t^i = 15\%$}       & $\chi=0$ & 1,236,177.94 & -10,609.31  & 53,561.03 & 67,670.50  & 97.48  \\
                                            & $\chi=1$ & 1,013,519.48 & -9,453.77   & 46,515.32  & 72,564.44 & 97.40 \\\hline           
\end{tabular}
}
\end{table}

The first and one of the most important results that we can confirm out-of-sample is that allowing the large GENCO offers to deviate from marginal costs makes the market marginal price increase. Notice that the mean day-ahead price offering at production cost would be 96.66 \euro{}/MWh.

Furthermore, for both the risk-neutral and risk-averse GENCOs, their profit increments are directly related to the price flexibility that we allow them to take. This fact makes our stochastic results previously exposed true also in a real market test. We can see how the total profit increment for the risk-neutral GENCO is higher than the one for the risk-averse producer for all cases of price flexibility levels.

Regarding the profit increment quantiles, i.e., on which values we expect the increment in profit to move on a daily basis; we can see that the range of values increases when also increasing the price flexibility level. For example, we can notice that for the 5\% flexibility level, a daily profit increase with the optimal strategy should be expected to be below 20,000 \euro{}.

In relation to the deviation of the profit increment when the optimal strategy is chosen, the risk-averse producer is not able to hold this variability as well as the risk-neutral GENCO. This fact can also be seen in the first quantile, which is generally lower in the risk-averse GENCO case, except for the 15\% flexibility level. This is due to some specific days where the risk-averse optimal strategy is worse than offering at marginal cost, which can now be checked in the graphical analysis.

\begin{figure}[H]
    \centering
    \begin{subfigure}{\textwidth}
    \centering
    \includegraphics[width=0.65\textwidth]{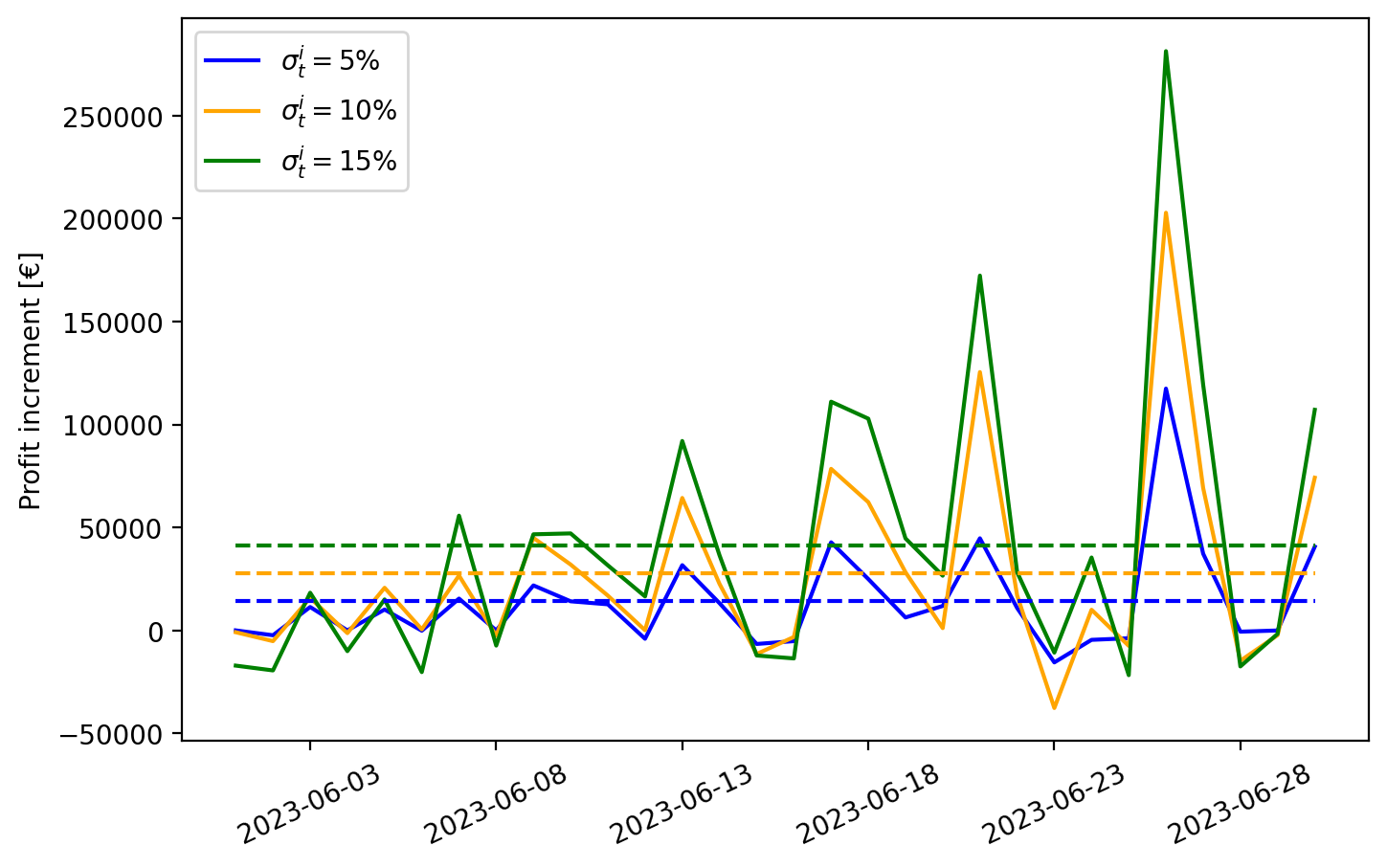}
    \end{subfigure}
    \hfill
    \begin{subfigure}{\textwidth}
    \centering
    \includegraphics[width=0.65\textwidth]{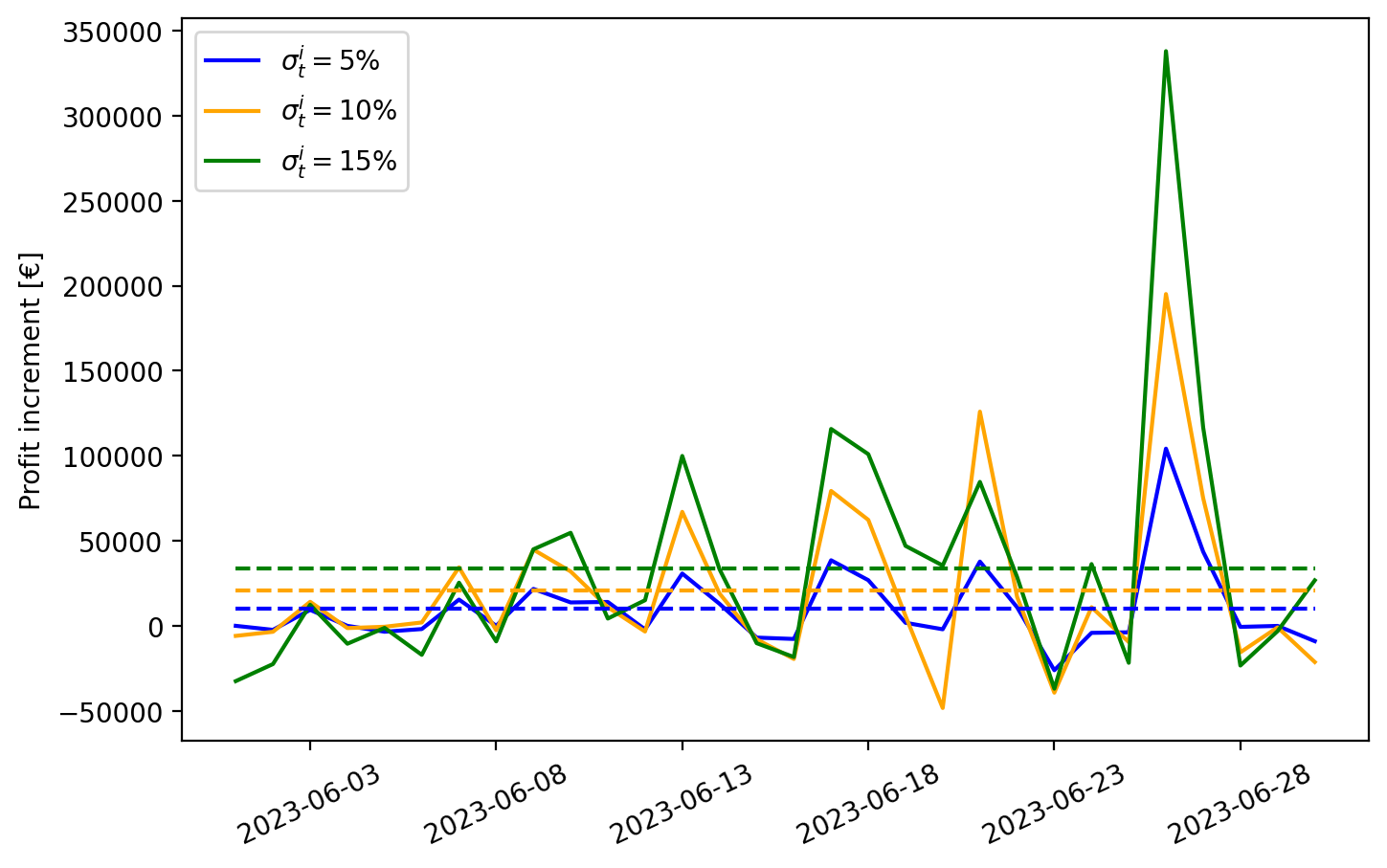}
    \end{subfigure}
    \caption{Daily out-of-sample profit increment for risk-neutral (up) and risk-averse (down) GENCO at different levels of price flexibility}
    \label{fig:oos_day_prof}
\end{figure}

Therefore, to finish this section, we will show graphically the daily GENCO profit increase in the test period for different levels of price flexibility and risk aversion concerning the true generating cost offering strategy. Figure \ref{fig:oos_day_prof} shows the daily out-of-sample profit increment for risk-neutral (up) and risk-averse (down) GENCO for different levels of price flexibility. Dashed lines represent the mean for each pricing strategy.

As can be seen, on most days, the profit increment is slightly higher for cases where the price flexibility is different from zero, for both levels of risk aversion. Also, on some specific days, this profit difference is increased on a higher scale, for instance, June 13, 17, 21, or 25. It is interesting to notice that, for both risk-aversion levels, the mean profit increment can rise to 50,000\euro, being slightly higher in the risk-neutral GENCO. Furthermore, it is interesting to notice how small flexibility levels help to increase the profit increment with respect to the cost-based strategy when the increment is negative.

\begin{figure}[H]
    \centering
    \begin{subfigure}{\textwidth}
    \centering
    \includegraphics[width=0.65\textwidth]{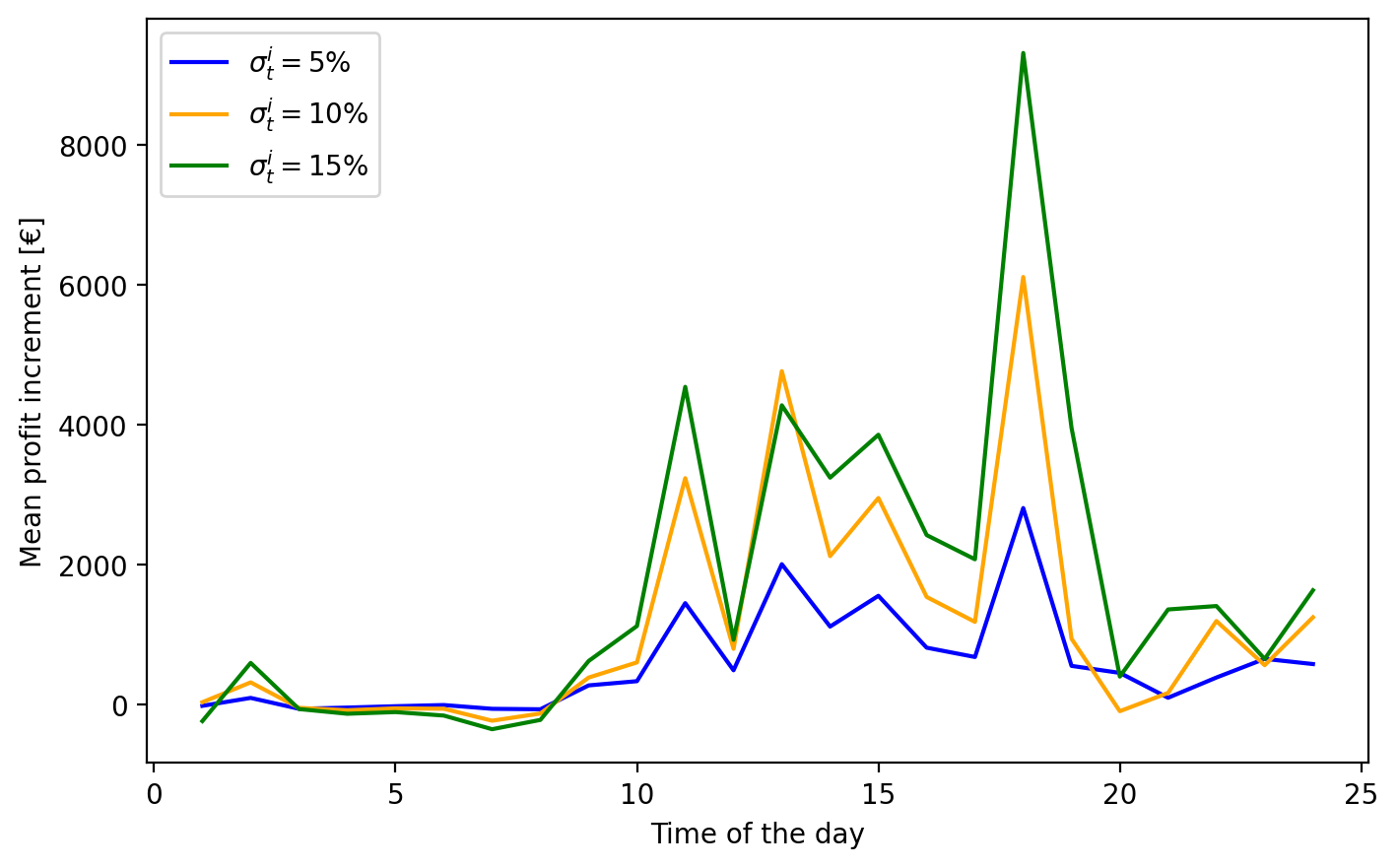}
    \end{subfigure}
    \hfill
    \begin{subfigure}{\textwidth}
    \centering
    \includegraphics[width=0.65\textwidth]{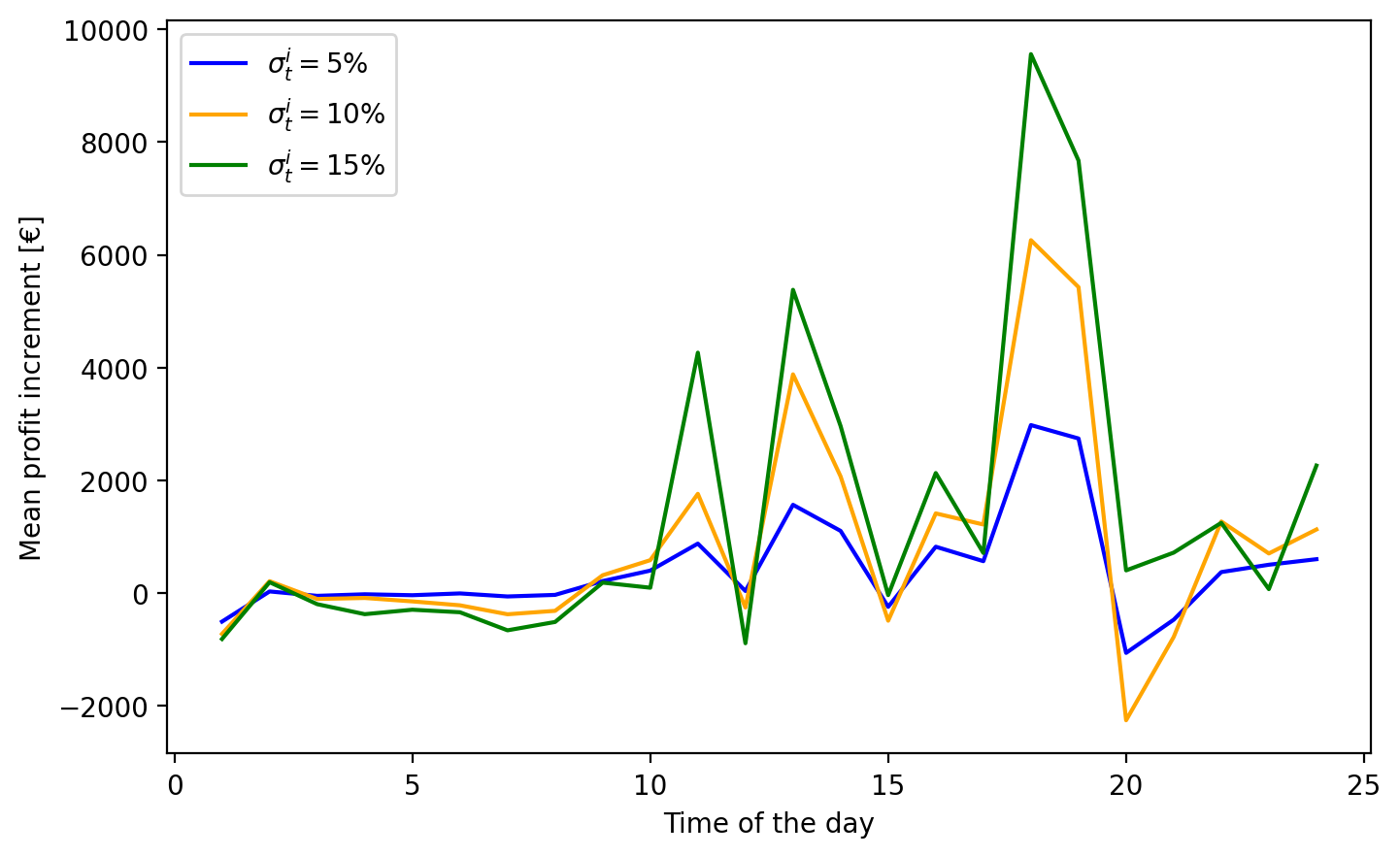}
    \end{subfigure}
    \caption{Hourly out-of-sample mean profit increment for risk-neutral (up) and risk-averse (down) GENCO at different levels of price flexibility}
    \label{fig:oos_hour_prof}
\end{figure}

Furthermore, we can appreciate more differences when we show the average increase in profits on an hourly basis (Figure \ref{fig:oos_hour_prof}). As in the case above, we represent the profit increment for several levels of price flexibility and risk aversion: risk-neutral (up) and risk-averse (down). We can see how the biggest amount of profit increment can be made after 8:00, and in some specific hours like 11:00 to 15:00, and 18:00 in a non-homogeneous way.

For the risk-neutral GENCO, we can notice how, on average, and in most of the hours of the day, a profit increment can be obtained by allowing price flexibility in her offering strategy. Besides, when the profit increases, it does it on a bigger scale when the maximum price flexibility is set. Furthermore, only in hours before 8:00, the profit increment is slightly below zero, i.e., a cost-offering strategy is better.

On the other hand, the behavior of the hourly profit increments is similar in the case of a risk-averse GENCO. However, there are more hours when the mean profit does not increase, such as 20:00. In the case where the profit suffers an increment, it does it proportionally to the different levels of price flexibility. Nevertheless, the case of a 5\% price flexibility allows the risk-averse GENCO to soften the decrease in profit.

These results show how the stochastic optimization model is optimistic about the profit increment. Nevertheless, the profit increments do occur, but at a lower scale than the reported stochastic model. This could be due to local limitations of the predictive model, as large variations in competitor block offers, which could change market outcomes, cannot be taken into account. However, the hourly behavior of the profit increment provides valuable insights and suggests that an hourly price flexibility strategy could be taken into account to further increase GENCO's profits.

To sum up, this out-of-sample validation approach has allowed us to confirm in a real-life context how allowing a GENCO to deviate from its true generating costs makes her expected profits increase. Nevertheless, we have seen how an excessive value of the price flexibility may create too much volatility in the results, in particular, in the risk-averse GENCO. Finally, we have learned that the highest increase in profit can be achieved at some specific hours of the day.

\section{Conclusions}
\label{sec:conclu}

In this work, we have dealt with the problem of finding an optimal offering strategy for a large GENCO. That is, knowing the amount of energy that the GENCO can produce and at what cost, setting the prices of the offered energy blocks for a profit-maximizing strategy.

We present a data-driven methodology where GENCO's supply curves and the ones from the rest of the competitors are optimally discretized. This discretization allows to get important insights from their offering strategy, reproduce the market clearing to compute the resulting market price, and ease an out-of-sample validation of the proposed stochastic optimization model.

The relationship between the hourly market marginal price and the block prices offered by the GENCO is modeled through a Distributional Neural Network approach. The advantage of this approach is to obtain an accurate but piece-wise linearizable model in its structure, from which different scenarios for the market outcomes can be sampled from their predicted distribution with a simple linear constraint. This predictive model is embedded into a two-stage stochastic optimization model which accounts for risk aversion to derive the optimal supply curve to submit to the day-ahead market.

After an in-depth analysis, stochastic results have shown how allowing the GENCO to deviate from her marginal costs offer, results in a marginal market price increase, which also increases her profits. Besides, there are differences in the pricing behavior of the GENCO depending on her risk aversion level. In general, a risk-neutral GENCO achieves to increase the marginal price to a higher extent than the risk-averse GENCO, but worsens the worst case profit scenarios.

One of the main novelties of this work is that the optimal offering strategy is tested out-of-sample. That is, we simulate the actual functioning of the market by combining the demand and the offers from the GENCO and her competitors. We show how the proposed optimization model achieves a marginal price increment and a profit increase for risk-averse and risk-neutral GENCOs when they deviate from marginal cost offerings. Besides, an hourly price flexibility strategy could further increase GENCO's profit. Furthermore, these results warn us about the importance of executing effective audits in markets with uniform pricing based on marginal technologies. Minor increments in the block prices may significantly increase the marginal price which will be transferred to the consumers.

Future work is focused not only on studying the price strategy for the GENCO but also on allowing flexibility on the offered energy quantity. This would add more complexity to the optimization problem and the prediction model. Furthermore, hourly-trained models could be fitted to better capture different market dynamics. \cblue{Besides, learning the functional relationship between the marginal price distribution and only the bids that are likely to be marginal could improve the performance of the methodology.} Finally, it would be possible to avoid the normality assumption by making use of a quantile neural network. In this case, we would need to estimate a set of quantiles that may act as scenarios in the stochastic optimization problem.

\section*{Acknowledgements}
The authors gratefully acknowledge the financial support from MCIN/AEI/10.13039/ 501100011033, project PID2020-116694GB-I00 and from the FPU grant (FPU20/00916).

\printbibliography

\end{document}